\documentclass{aa}  

\usepackage{graphicx}
\usepackage{txfonts}
\usepackage{lipsum}
\usepackage{siunitx}
\usepackage{subcaption}         
\usepackage{lscape}             
\usepackage{placeins}           
\usepackage{multirow}
\usepackage{tablefootnote}
\usepackage{hyperref}

\usepackage{xcolor}                          

\begin{document}

   \title{ALMA high-resolution observations of Betelgeuse: Persistent structure spanning the inner atmosphere}

\author{W.R.F. Dent\inst{1} 
   \and A.M.S. Richards\inst{2} 
   \and G.M. Harper\inst{3} 
   \and L.D. Matthews\inst{4} 
   \and E. O'Gorman\inst{5} 
}        

   \institute{ESO/ALMA, Karl-Schwarzschild-Straße 2, 85748 Garching bei München, Germany.   billdent42@gmail.com\\
    \and Jodrell Bank Centre for Astrophysics, University of Manchester, Manchester, UK.  a.m.s.richards@manchester.ac.uk    
    \and Center for Astrophysics and Space Astronomy, University of Colorado, 389 UCB, Boulder, CO 80309, USA.  Graham.Harper@colorado.edu
    \and  Massachusetts Institute of Technology Haystack Observatory, 99 Millstone Road, Westford, MA 01886 USA.  lmatthew@mit.edu
    \and Dublin Institute for Advanced Studies, 31 Fitzwilliam Place, Dublin 2, Ireland
eamon.ogorman@gmail.com}

   \date{}
\authorrunning{W.R.F. Dent et al.}

  \abstract
  {The extended atmosphere of red supergiants (RSGs) forms an important link in the process of mass loss and the subsequent enrichment of the interstellar medium. Large-scale convection is thought to play a significant role, which is likely to result in irregularities in the surface.}
  {High-resolution, high-contrast sub-millimetre images of Betelgeuse - one of the closest RSGs - are used to probe the structure and temporal stability of the inner 1-2$R_\star$ of its atmosphere.}
  {Using ALMA in the longest baseline configuration, continuum emission and lines of SiO and CO and their isotopomers were observed at 0.6-1.4~mm, giving beamwidths down to 7~mas at the shortest wavelengths. These were compared with a similar observation taken at 0.9~mm approximately 7 years earlier.}
  {Continuum emission arises mostly from an optically-thick sub-millimetre photosphere of radius 1.1-1.3~$R_\star$ with a relatively constant temperature of $\sim$2300~K, but with two hotter patches to the NE and SW. The brightest of these has a temperature enhancement of $\sim$800K, and its location and intensity appear relatively unchanged since the 2015 observation. The sub-millimetre photosphere shows deviations of up to $\pm$ 6\% in radius, with weaker continuum extending out to $\sim$2.5$R_\star$ - similar to the extent of clumpy emission in SiO and CO.}
 {The hot regions of gas and deviations from radial symmetry are thought to be associated with active shocks driven by underlying convective cells, although their lifetimes appear to be longer than model predictions. The hotter regions lie near the proposed poles of the star, which might suggest enhanced and relatively stable polar convection. The present data show no clear evidence for stellar rotation in the extended line emission or absorption against the photosphere, although the structure of the gas emission has changed significantly since 2015.}

   \keywords{Stars: --
                supergiants
               }

   \maketitle
\nolinenumbers

\section{Introduction} \label{sec:intro}

Betelgeuse (HD39801, $\alpha$~Ori) is one of the two closest red supergiant (RSG) stars and, as the archetype, it has been the subject of intense study from radio through to X-ray wavelengths. Its relatively large photospheric angular
    diameter of $\simeq 43$\,mas  \citep{Montarges2021} also enables it to be resolved with current instrumentation. There remains some uncertainty in the distance, primarily due to parallax errors arising from photocentre shifts caused by its large angular diameter and photospheric inhomogeneities \citep{Macleod2025}. However, here we adopt the astrometric estimate of $172^{+13}_{-11}$~pc derived by \cite{Macleod2025}, which is consistent with the independent estimate of $168^{+27}_{-15}$~pc based on the modelling of astroseismology data \citep{Joyce2020}.

There is growing evidence that the optical photosphere and inner atmospheres of RSGs such as Betelgeuse are non-uniform.  Early \textit{Hubble} Space Telescope (HST) 
UV images showed a 100~mas diameter chromosphere with a bright spot offset $\sim$20~mas from the centre \citep{Gilliland1996}. Non-redundant masking techniques\footnote{Effectively an interferometer with apertures distributed over a primary mirror} at optical wavelengths with a maximum baseline of $\sim$6~M${\lambda}$ \citep{Buscher1990, Tuthill1997} showed bright spots or patches on the photosphere. 
Interferometry at near-infrared and optical wavelengths using three or four separate telescopes can provide maximum baselines of $\sim$25-60~M${\lambda}$, although limited {\it uv} coverage requires parametric model fitting to infer  the surface structure \citep{Young2000, Haubois2009, Montarges2016, Ohnaka2018, Drevon2024}. Inversion modelling of optical line polarisation has also been used to derive the surface topology in terms of spherical harmonics \citep{LopezAriste2018}. The Atacama Large Millimeter/submillimeter Array (ALMA) has baselines of up to 16~km ($\sim$27~M${\lambda}$ at 850$\mu$m) with typically $\sim$40 dishes, and the resulting relatively dense {\it uv} coverage can provide high fidelity images with resolutions of $\sim$15~mas in the continuum at this wavelength. ALMA observations at 850$\mu$m in 2015 showed evidence of a bright region offset to the NE of the stellar centre \citep{OGorman2017}.

\cite{Schwarzschild1975} postulated that RSGs may contain a few large convective cells to explain some of the variability on timescales of a few hundred days. This was reinforced by hydrodynamic models, showing that the combination of convective flows and non-radial stellar pulsations can generate surface structure on scales of $\sim$10\% of the stellar radius, $R_\star$ \citep{Freytag2002}. These may merge to form a few large-scale hot regions, particularly when observed at relatively low resolution \citep{Freytag2002, Ahmad2023,  Freytag2024, Ma2024}.
Specific models of Betelgeuse suggest that these may be partially responsible for its semi-regular variability \citep{Pilate2024}. A combination of convective cells with stellar pulsation is predicted to drive radiative shocks and outflows into the lower atmosphere \citep{Chiavassa2024}, which cool to allow the formation of molecules such as H$_2$O and SiO \citep{Tsuji2000}, and eventually dust \citep{Verhoelst2006}. This may explain the molecular shells (or MOLspheres) and plumes traced by SiO and CO at a few stellar radii \citep{Kervella2018, Tsuji2000}. 

The intense study of Betelgeuse has revealed additional asymmetric and time-dependent characteristics. For a few months in early 2020, the visible brightness dropped by 50\% (the so-called `Great Dimming'), which was associated with a localised cool photospheric patch, possibly combined with an absorbing foreground dust clump \citep{Montarges2021}. More recently, some of the periodic variability and radial velocity changes have been attributed to a 
nearby low-mass companion star with a semi-major orbital axis of 2.3$R_\star$ \citep{Goldberg2024, Macleod2025, Montarges2026}. In summary, published observations suggest that Betelgeuse is not a smooth symmetric well-behaved star, and high-resolution, high-fidelity imaging is needed to unambiguously delineate its structure.

In this work, we present observations of Betelgeuse in lines and continuum at ALMA Bands 6, 7, and 8 (in the range 215 to 492~GHz) taken in 2023. These were obtained when the observatory was in its longest baseline configuration, achieving resolutions down to $~\sim$7~mas (17\% of the stellar diameter) and imaging dynamic range of $>$100 in the highest frequency band. Section 3 presents the results in both continuum and lines, along with an axisymmetric model. In Section 4, we discuss possible origins of the asymmetric structures, comparing the data with those taken 7 years ago and with the location of the proposed companion.

\section{Observations} \label{sec:observations}

Betelgeuse was observed 
using ALMA in its most extended configuration (maximum baseline 16~km) in
five executions during August 2023. Bands 6, 7, and 8 were observed with spectral windows configured to provide the maximum continuum bandwidth of 7.5~GHz, as well as simultaneously covering multiple
transitions of SiO and CO and their isotopes, and other species such as the atomic [{\mbox{C{\sc i}}}] transition in Band 8 and the Rydberg lines in Band 6.  The details of the observations, spectral setups, and observing conditions are given in Table~\ref{tab:obs_details}.

The on-source integration time per execution was a total of 45~minutes. Standard phase referencing was used, cycling every 80--90 seconds to the phase calibrator J0552+0313 (at 4.2$^{\circ}$ separation) in Bands 6 and 7, and J0522-3627 (separation 5.6$^{\circ}$) in Band 8. A check source (J0558+0044, separation 2.8$^{\circ}$ from the phase calibrator at Bands 6 and 7, and J0552+0313, separation 6.6$^{\circ}$ at Band 8) was also observed two or three times during each execution and used to estimate astrometric accuracy. The bandpass was calibrated using J0510+1800 at Bands 6 and 7 and J0532+0732 at Band 8, and these were also used as secondary flux calibrators.
Observations used a target proper motion of (+27.5, +11.3~mas/yr), a parallax of 6.56~mas, and the observing frequency was corrected to the source velocity of ${\rm v}_\star$ = +4.9~km~s$^{-1}$ in the local standard of rest frame (LSR) \citep{vanLeeuwen2007}. 
For each band, the standard calibrated measurement sets of visibility data from the ALMA \textsc{casa}\footnote{https://casa.nrao.edu \citep{CASA2022}.} pipeline were concatenated, and the line-free channels were used to make continuum images and provide a model for subsequent phase self-calibration and additional amplitude self-calibration in Band 7. These solutions were applied to all spectral channels.

For the restoring beams in continuum, we employed different {\it uv} weightings for different purposes. Briggs weighting with robustness parameter R=0.5 was initially used, being the default for ALMA as it gives a balance between resolution and sensitivity, and in this case achieved image dynamic ranges of 2700, 1600, and 170 at Bands 6 through 8. For this and all continuum images, we included a linear spectral dependence of flux and used multi-scale components with multi-term multi-frequency synthesis (\texttt{mtmfs})\footnote{https://casadocs.readthedocs.io/en/stable/notebooks/synthesis\_imaging.html}.
The very high S/N ratio and relatively dense {\it uv} coverage enabled us to obtain up to $\sim$40\% higher resolution by increasing the weighting of the longer baselines using the so-called `super-uniform' option in \textsc{casa} \texttt{tclean}$^{3,}$~\footnote{'Super-uniform' gives a similar resolution to uniform weighting, except that it reduces inner sidelobes and increases sensitivity by averaging the {\it uv} weighting over small groups of cells (in this case by npixels=10). Particularly useful for long-baseline configurations where there are relatively large gaps between the distant antennas.}. This resulted in the restoring beams listed in Table~\ref{tab:obs_details}. Although it results in higher resolution, this option increases the image noise level by $\sim$50\% compared to Briggs R=0.5. It also up-weights the more sparsely distributed long-baseline antennas - potentially increasing imaging artefacts - and so the effects on the data were investigated by comparing with images using R=0.5 and observation simulations (Appendix~\ref{sec:appendix_sim}).  These comparisons and the good agreement between images in the three frequency bands (with their correspondingly different {\it uv} coverages) show that such artefacts are small and do not contribute significantly to the structure in the final results.

We made spectral image cubes after continuum subtraction, using Briggs R=0.5 weighting. Line and continuum images were also made with other weightings, including downweighting the longest baselines for images optimised for larger-scale sensitivity.
The maximum recoverable scales range from 220~mas in Band 6, down to $\sim$110~mas in Band 8. The absolute flux calibration accuracy (including uncertainty in the flux calibrators and from the flux transfer to the science target) is assumed to be $\sim$ 5\% at Band 6, 7\% at Band 7, and 10\% at Band 8\footnote{ALMA Cycle 10 Technical Handbook.}. The astrometric accuracy of the data in Band 7 was estimated to be $\sim$1~mas (see Sect.~\ref{sec:cont_results}).

\begin{table*}
	\caption{Observing details. }
    	\label{tab:obs_details}
	\centering

    \begin{tabular}{c c c c c c c c}
		\hline\hline
        
Band : SPW & Centre freq. & Dates & PWV & Phase noise & Beam & Max. recoverable & Continuum \\
 & (GHz)$^{(1)}$ & (Aug. 2023) & (mm) & (rms, degr.)$^{(2)}$ & (mas, PA)$^{(3)}$ & scale (mas)  & rms ($\mu$Jy)$^{(3)}$\\
 \hline
6 : 1  & 214.800 & \multirow{4}{3em} {03, 27} & \multirow{4}{3em}{0.3, 1.1} & \multirow{4}{3em}{26, 32} & \multirow{4}{3em}{19.6~$\times$~14.1 $@ 47^{\circ}$} & \multirow{4}{3em}{220} & \multirow{4}{3em}{33}\\ 
6 : 2  & 217.105 &  &  &  &  &\\
6 : 3  & 230.300 &  &  &  &  &\\
6 : 4   & 232.000 &  &   &  & &\\
\hline
7 : 1 & 331.100 & \multirow{4}{3em}{03, 04} & \multirow{4}{5em}{0.32, 0.99} & \multirow{4}{3em}{25, 50} & \multirow{4}{3em}{10.7~$\times$~9.5 $@ 19^{\circ}$} & \multirow{4}{3em}{140}& \multirow{4}{3em}{39}\\

7 : 2 & 332.550 &  &  &  &  &\\
7 : 3 & 343.200 &  &  &  &  &\\
7 : 4 & 345.100 &  &  &  &  &\\
\hline
8 : 1 & 478.200 & \multirow{4}{3em}{01} & \multirow{4}{3em}{0.38} & \multirow{4}{3em}{30} & \multirow{4}{3em}{7.7~$\times$~6.6 $@ 2^{\circ}$}& \multirow{4}{3em}{110} & \multirow{4}{3em}{234} \\

8 : 2 & 480.200 &  &  &  &  &\\
8 : 3 & 490.300 &  &  & &    &\\
8 : 4 & 492.200 &  &  &  &  &\\

\hline

\end{tabular}

 (1) Centre frequency of each spectral window (SPW) is given in the adopted rest frame of the source 
 (${\rm v}_{\mathrm{LSR}}$~=~4.9~km~s$^{-1}$). Each has a bandwidth of 1.875~GHz, a channel spacing of 0.976~MHz, and an effective spectral resolution (using Hanning spectral smoothing) of 1.95~MHz;
 (2) Phase noise represents the atmospheric phase stability, and is the scan-to-scan phase rms (in degrees), measured on the phase calibrator on the longest baselines;
 (3) Beam dimensions, position angle (PA), and rms noise level are given for the continuum, combining spectral windows and datasets using \texttt{mtmfs} in \textsc{casa} \texttt{tclean} with super-uniform weighting$^{2,4}$.
 
\end{table*}

\section{Results}

\subsection{Continuum}
\label{sec:cont_results}

\begin{figure}

\begin{center}
	\includegraphics[width=7.1cm,angle=-90]{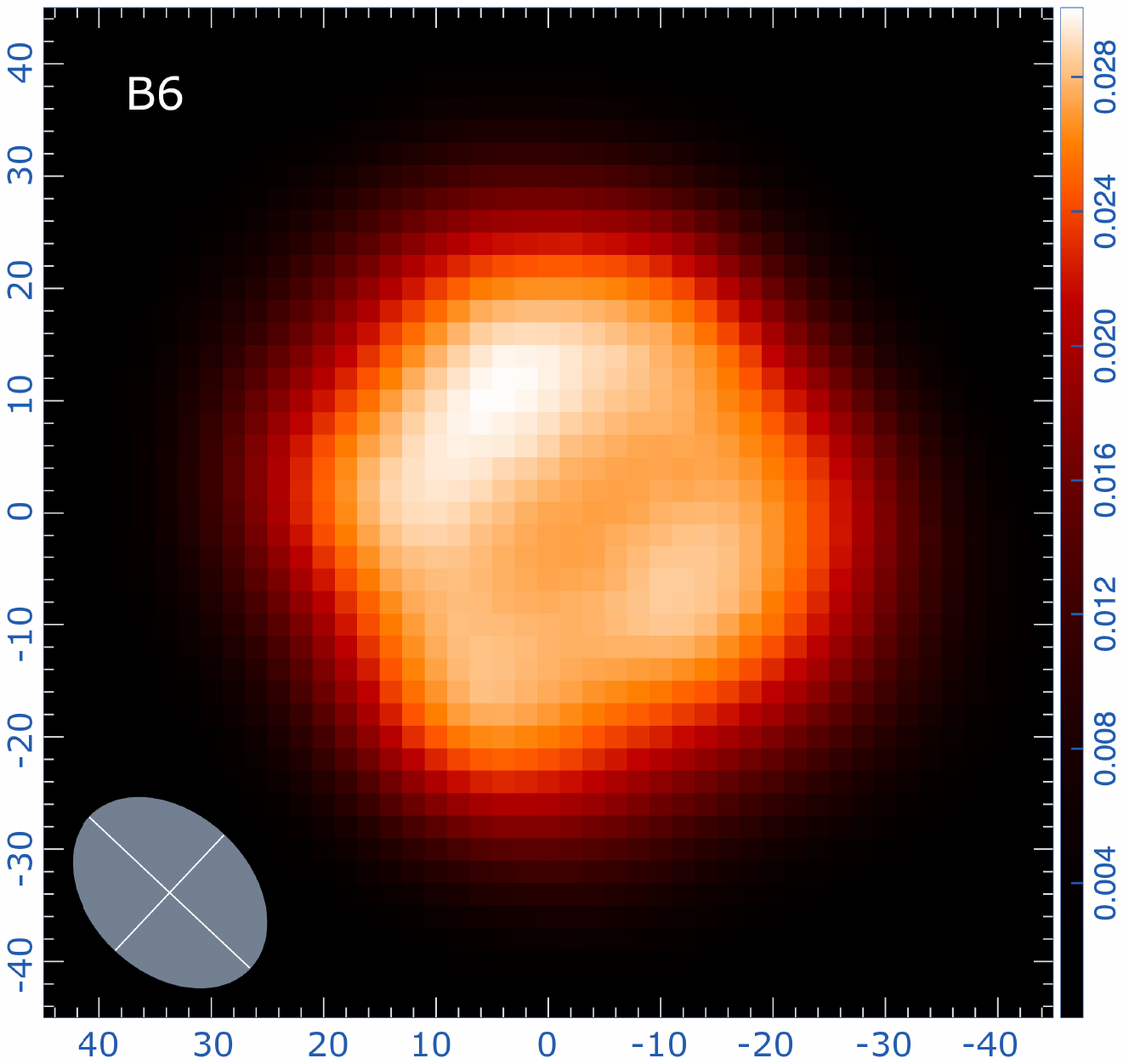}
 	\includegraphics[width=7.1cm,angle=-90]{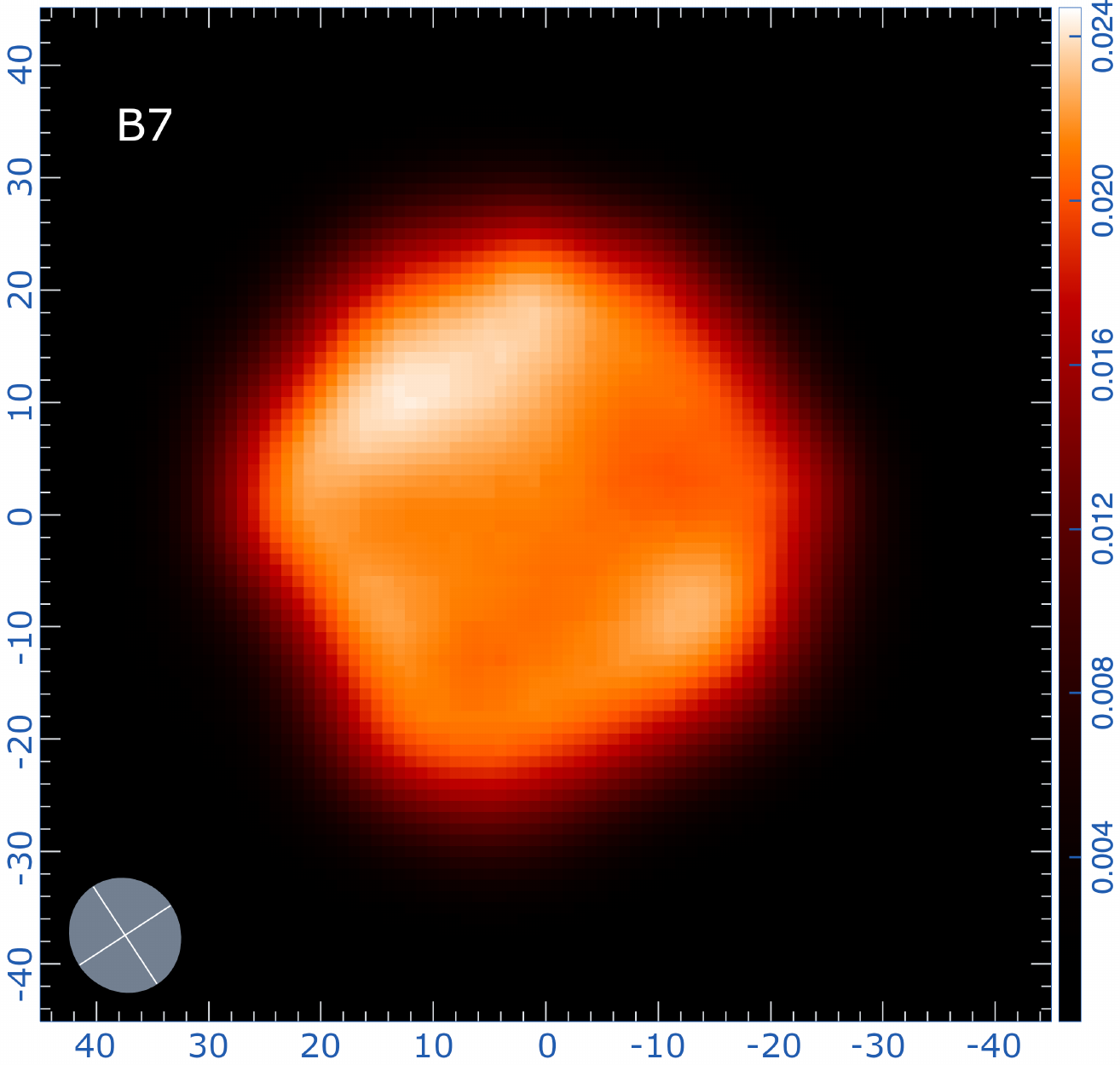}
     \includegraphics[width=7.1cm,angle=-90]{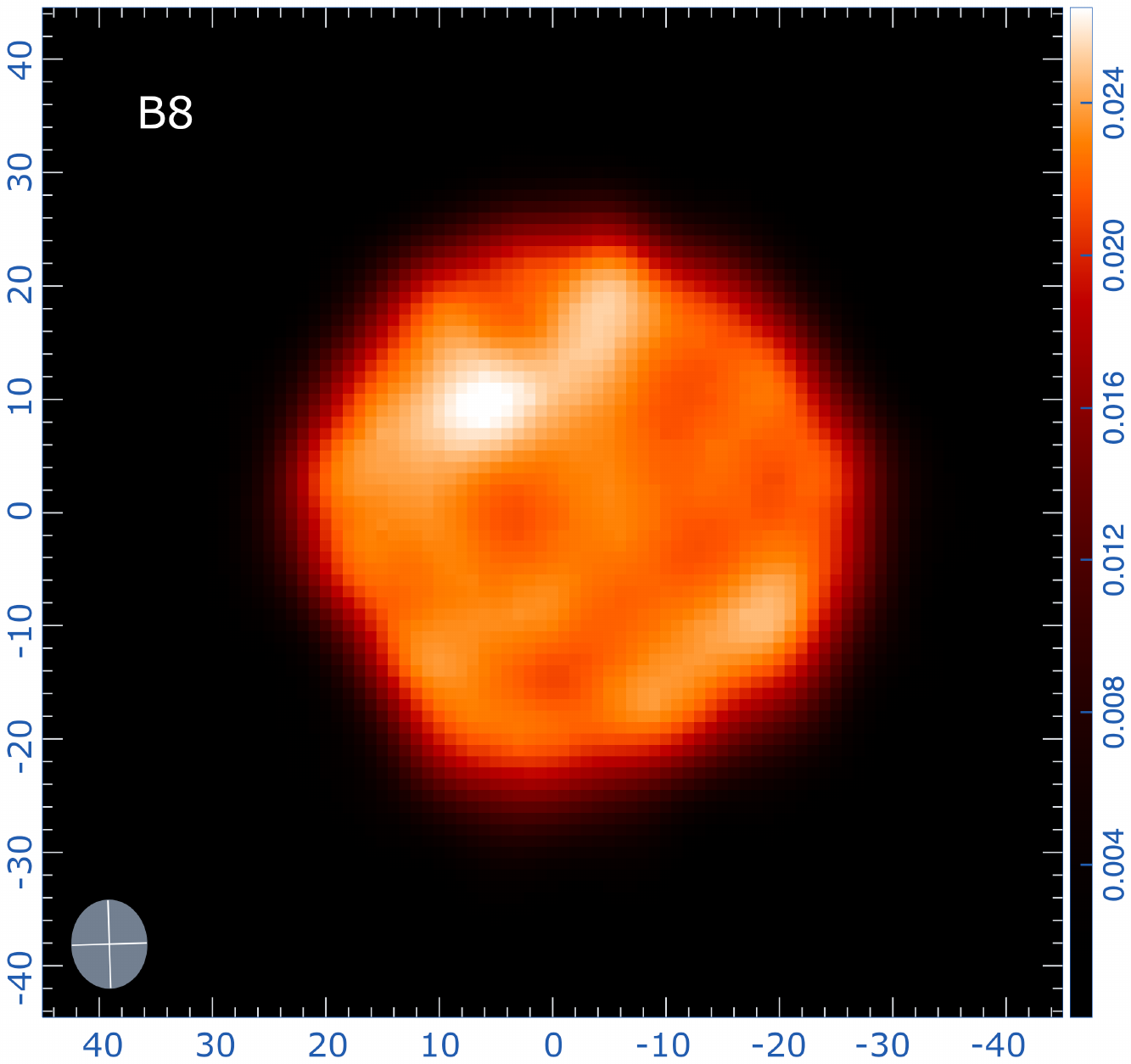}
    \caption{Continuum images of Betelgeuse in Bands 6, 7, and 8 restored with super-uniform {\it uv} weighting. Colour tables are illustrated by the colour bars in units of Jy/beam, and use square power law stretches to show the structure in the inner atmosphere. Field of view is 90~mas, and the beam dimensions are shown lower left, with the smallest beam of 7.7$\times$6.6~mas at Band 8. Positional coordinates are equatorial offsets in mas.
    }
	\label{fig:contin_images}
    \end{center}
\end{figure}

The restored high-resolution continuum images in the three bands are shown in Fig~\ref{fig:contin_images}. Integrated fluxes for Bands 6, 7, and 8 are 0.28$\pm$0.014, 0.51$\pm$0.04, and 0.90$\pm$0.09~Jy at mean frequencies of 223.55, 337.99, and 485.22~GHz, respectively. These fluxes were obtained from photometry on restored images with a 120~mas diameter aperture. The main contribution to the errors is the absolute calibration uncertainty. They are consistent with the measurements of 0.55~Jy in Band 7 in 2015 \citep{OGorman2017} and 0.29~Jy at 230~GHz using CARMA in 2007-2009 \citep{OGorman2012}.  The stellar position was estimated by fitting the centroid of the emission below 80\% of the peak before self-calibration to minimise skewing the result because of the asymmetrical emission over the surface. In the combined Band 7 dataset, the measured stellar position was 5$^h$55$^m$10.3438$^s$, 7$^{\circ}$24'25.641" (ICRS, uncorrected for parallax), at date MJD 60160. This was found both by measuring the image centroid and by fitting visibilities in the {\it uv}-plane. The Band 7 data were estimated to have an astrometric accuracy of $\sim$1~mas, the best of the three bands. This was derived from measurements of the nearby check source scaled by the relative separation, as well as from the phase noise and a comparison of the two independent executions (as used by \citealt{Gottlieb2022}). The measured astrometric equatorial offsets of the image centroid of the photosphere relative to this in Bands 6 and 8 were (RA,decl.) = ($-$3.6,$-$1.1) and ($-$2.5,$-$4.2)~mas respectively, or a separation of 21\% and 54\% of the average full width at half maximum (FWHM) beamsize in these bands. These astrometric errors are typical of ALMA in the longest baseline configuration$^5$, and are likely due to uncertainties from the phase reference calibration (particularly at Band 8), the larger beam at Band 6, and the asymmetric shape of the star.

\subsubsection{Structure on the surface}
\label{sec:surface_structure}
The images in all three bands clearly show an emission peak offset by $\sim$12~mas to the NE of the centre; this is close to that noted in the Band 7 data taken with a similar resolution in 2015 \citep{OGorman2017}. Additionally, there is evidence of a fainter peak $\sim$18~mas to the SW - most clearly seen in the higher resolution images in Bands 7 and 8, although this was less distinctly a separate peak in the 2015 data.

\begin{figure}

\begin{center}
     \includegraphics[width=8cm,angle=0]{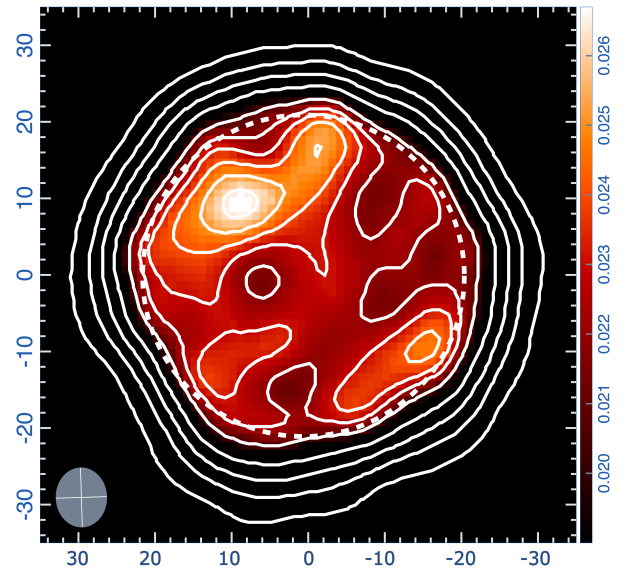}
    \caption{Contour map of Betelgeuse at Band 8 restored with super-uniform {\it uv} weighting. The nominal stellar size is illustrated by the 42~mas diameter dashed circle. Contour levels are 4,8,12,16,20,22,23,24,25,26~mJy/beam, where 1~mJy/beam  = 5$\sigma$. A linear colour table stretch from 19-27~mJy/beam is used to show the structure on the optically-thick surface. Field of view is 70~mas, and the beam shape is shown lower left. Positional coordinates on this and other images are equatorial offsets from stellar centroid in mas.
    }
	\label{fig:contin_image_B8}
    \end{center}
\end{figure}

The detailed structure seen in Fig~\ref{fig:contin_images} is consistent in the three bands after considering the different beam size, indicating that it is unlikely to be caused by imaging artefacts or random (thermal) noise. Comparison of the observations with an axisymmetric model confirms this (see Appendix~\ref{sec:appendix_sim}). In the Band 8 image (see also Fig.~\ref{fig:contin_image_B8}), the NE peak has a maximum flux/beam of 26.8~mJy, with a contrast over the median flux in a 42~mas centred aperture of 19$\sigma$. The emission to the SE has a maximum at Band 8 of 24.7~mJy, with a contrast of $\sim$11$\sigma$. There is also possible evidence of a central belt of enhanced emission at a position angle of $\sim$120$^{\circ}$ (similar to the extension of the NE and SW peaks), although the contrast against the regions either side is only $\sim$3$-$4$\sigma$, and this would need confirmation with higher resolution and S/N ratio.

As this emission is optically thick and the vertical temperature gradient in the emitting region is relatively shallow \citep{OGorman2017}, the observed flux can be used to estimate the local gas temperature $T_{\rm g}$ of what is effectively a millimetre or sub-millimetre photosphere\footnote{For brevity, we refer to this as the sub-millimetre photosphere} \citep{Reid1997}. In the restored images in Fig.~\ref{fig:contin_images}, the median brightness temperature $T_{\rm b}$ in the central 42~mas diameter region covering the stellar disk (outside the NE and SW peaks) is $\sim$2280~K in both Band 7 and 8, and $\sim$2400~K in Band 6 (although the latter suffers from worse resolution). Note that the certainty in the absolute values is limited by the flux calibration accuracy, but the relative calibration accuracy across the images is considered accurate within the rms noise (Table~\ref{tab:obs_details}; see also Appendix~\ref{sec:appendix_sim}). In these restored images, the maximum temperatures in the NE bright region are 2650, 2585, and 2734~K in Bands 6, 7, and 8. However, as this peak is only marginally resolved (even at Band 8), these are lower limits to $T_{\rm b}$, and a better estimate can be obtained by directly fitting the $\it uv$ visibilities rather than the restored image data. In Bands 6 and 7 we used a two-component model consisting of a uniform elliptical disk plus a 2-d Gaussian to fit the NE peak. The higher resolution at Band 8 shows that the northern patch breaks into two components, which we could fit with a Gaussian and a point source. The results are given in Table~\ref{tab:uv_fits}. Note that the listed position offsets are with respect to the main disk, so do not include the absolute astrometric errors noted in Sect.~\ref{sec:cont_results}. Also, the fits did not include the additional SW peak and extended emission (hence the total fluxes are slightly lower than the photometry in Sect.~\ref{sec:cont_results}). The positional errors are from the \texttt{uvmultifit} and do not include uncertainty due to source irregularity or over-simplification of the model. These fits to the visibilities indicate that the NE component has a temperature enhancement of between 500 and 800~K above that of the main sub-millimetre photosphere, and the region is elongated at PA$\sim$110$^{\circ}$. 

\begin{table*}
	\centering
	\caption{Best fit models for the star and NE region. }
	\label{tab:uv_fits}
	\begin{tabular}{lccccccc}
		\hline
&RA offset & Dec offset & Flux & Major axis & Axial ratio & PA & $T_{\rm b}$\\
&(mas) & (mas) & (mJy) & (mas) & & (deg) & (K)\\
\hline
{      }\underline{Band 6 } (223.55~GHz) &&&&&&&\\
Disk & 0 & 0 & 255.0$\pm0.2$ & 62.14$\pm0.02$ & 0.99$\pm$0.01 & 118$\pm1$ & 2342$\pm117$ \\
Gaussian & 11.0$\pm0.1$ & 19.0$\pm0.1$ & 19.2$\pm0.2$ & 34.8$\pm0.1$ & 1 & - & 546$\pm19$ \\
\hline
{      }\underline{Band 7 } (337.99~GHz)&&&& &&&\\
Disk & 0 & 0 & 486.1$\pm0.2$ & 57.74$\pm0.01$ & 0.98$\pm$0.01 & 32$\pm0.6$ & 2294$\pm115$ \\
Gaussian & 8.23$\pm0.07$ & 11.6$\pm0.14$ & 11.3$\pm0.1$ & 21.8$\pm0.2$ & 0.32$\pm0.01$ & 109$\pm0.4$ & 791$\pm36$ \\
\hline
{      }\underline{Band 8 } (485.22~GHz) &&&&&&&\\
Disk & 0 & 0 & 875.4$\pm0.9$ & 54.30$\pm0.04$ & 0.98$\pm$0.01 & 33$\pm1$ & 2266$\pm159$ \\
Gaussian & 10.1$\pm0.2$ & 9.2 $\pm0.3$ & 9.9$\pm0.7$ & 10.9$\pm1.0$ & 0.86$\pm0.05$ & 111$\pm0.4$ & 504$\pm75$ \\
Point source & -2.1$\pm0.4$ & 17.7$\pm0.4$ & 2.3$\pm0.2$ & - & - & - & - \\
\hline
\end{tabular}

\end{table*}

\begin{figure}
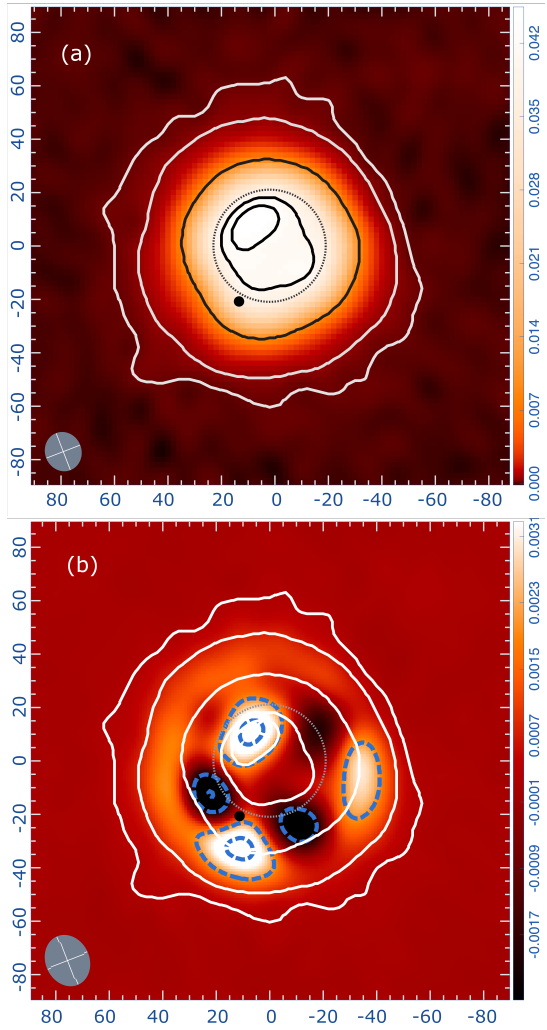

\begin{center}
    \includegraphics[width=7.1cm,angle=0]{Fig3a_BetelB7_3_cont_hi.pdf}
    \includegraphics[width=7.1cm,angle=0]{Fig3b_BetelB7_3_residuals_hi.pdf}
    \caption{(a) Band 7 image showing the continuum emission extending beyond the stellar photosphere (illustrated by the 42~mas diameter dotted circle). The image was restored using Briggs weighting with R=0.5, and is shown with a logarithmic colour table and contours levels (starting at 5$\sigma$) of 0.13,0.5,10,40,43~mJy/beam.  (b) Residuals after subtracting a symmetric model for the inner atmosphere (see text for details). Blue contours are $\pm$4,$\pm$2~mJy/beam. Field of view is 180~mas. Location of the proposed companion at the epoch of observations is shown by the black dot (see Sect.~\ref{sec:discuss-companion}). Colour bars to the right give the flux scale in Jy/beam.
    }
	\label{fig:B7_cont_R0p5}
    \end{center}
\end{figure}

The fits to visibility data yield a sub-millimetre photosphere diameter of 54.3~mas at Band 8, with the angular size increasing towards lower frequency bands (Table~\ref{tab:uv_fits}).
This results from the frequency dependence of the free-free optical depth, $\tau_{\nu} \propto \nu^{-2.1}$, and thus a larger $\tau_{\nu}=1$ surface at the lower frequencies \citep{Lim1998, Harper2001}, which continues towards Jansky Very Large Array (VLA) observations at cm wavelengths (\citealt{Matthews2022, Matthews2026}). A detailed model of this behaviour is described in Sect.~\ref{sec:radial_structure}. In addition to the bright sub-millimetre photosphere, fainter continuum emission is detected out to $\sim60$~mas, illustrated in the Band 7 image in Fig.~\ref{fig:B7_cont_R0p5}. Although this large-scale structure appears at first sight to be relatively smooth and symmetric (similar to the emission at 7~mm and 1.3~cm out to 4-5$R_\star$ \cite{Matthews2022}), the residuals after subtracting a symmetric model reveal variations in extended emission at a level of $\sim\pm$~10\% (Fig.~\ref{fig:B7_cont_R0p5}, lower panel). As the emission extending beyond the photospheric radius is likely to be optically thin (suggested by the spectral index maps; see Sect.~\ref{sec:radial_structure}), then this residual structure could be due to local variations in temperature and/or density.

\subsubsection{Spectral index and axisymmetric model}
\label{sec:radial_structure}

\begin{figure}
\begin{center}
     \includegraphics[width=8.cm,angle=0]{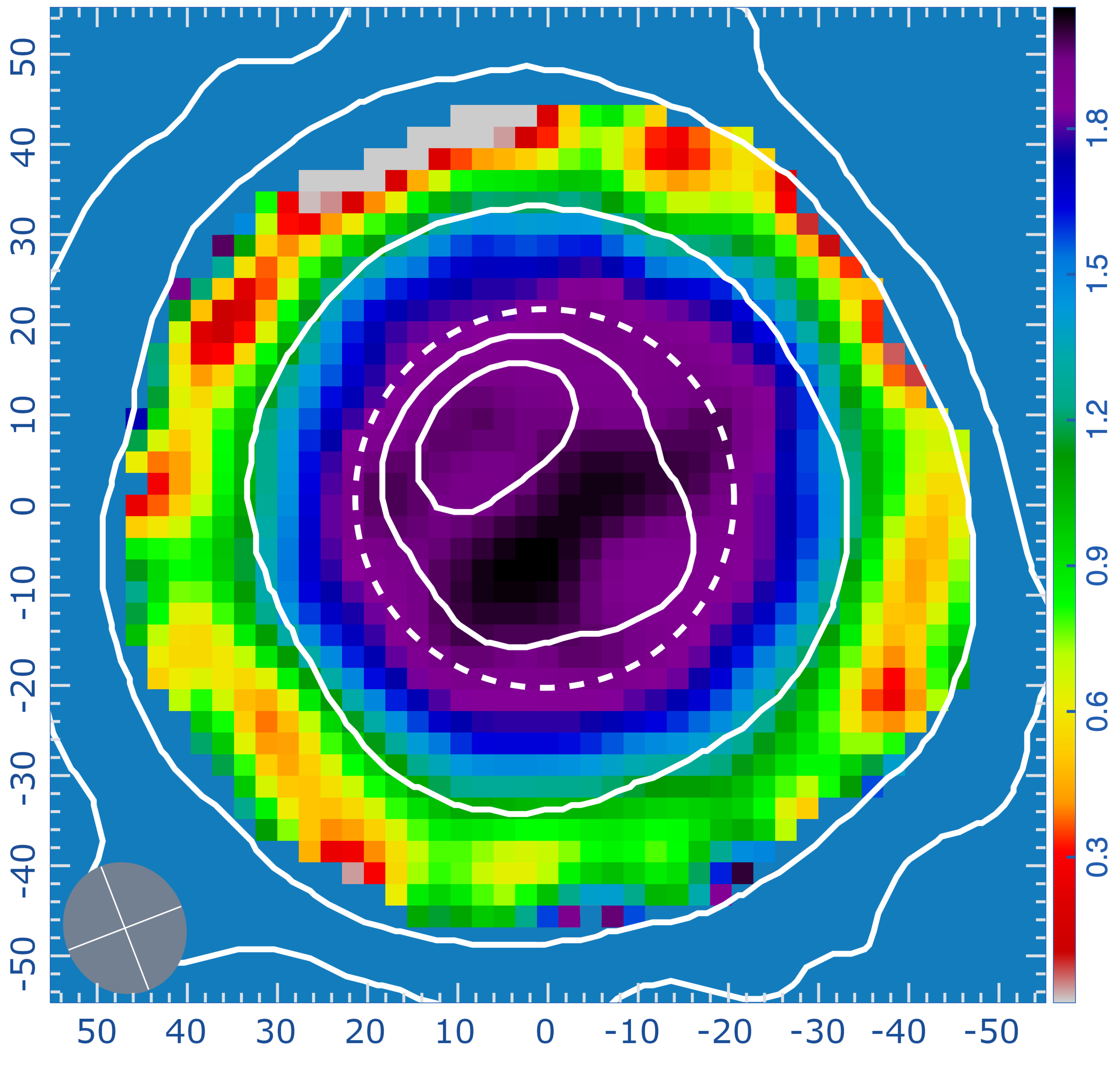}
    \includegraphics[width=9.cm,angle=0,trim=1cm 0cm 1cm 0cm,clip]{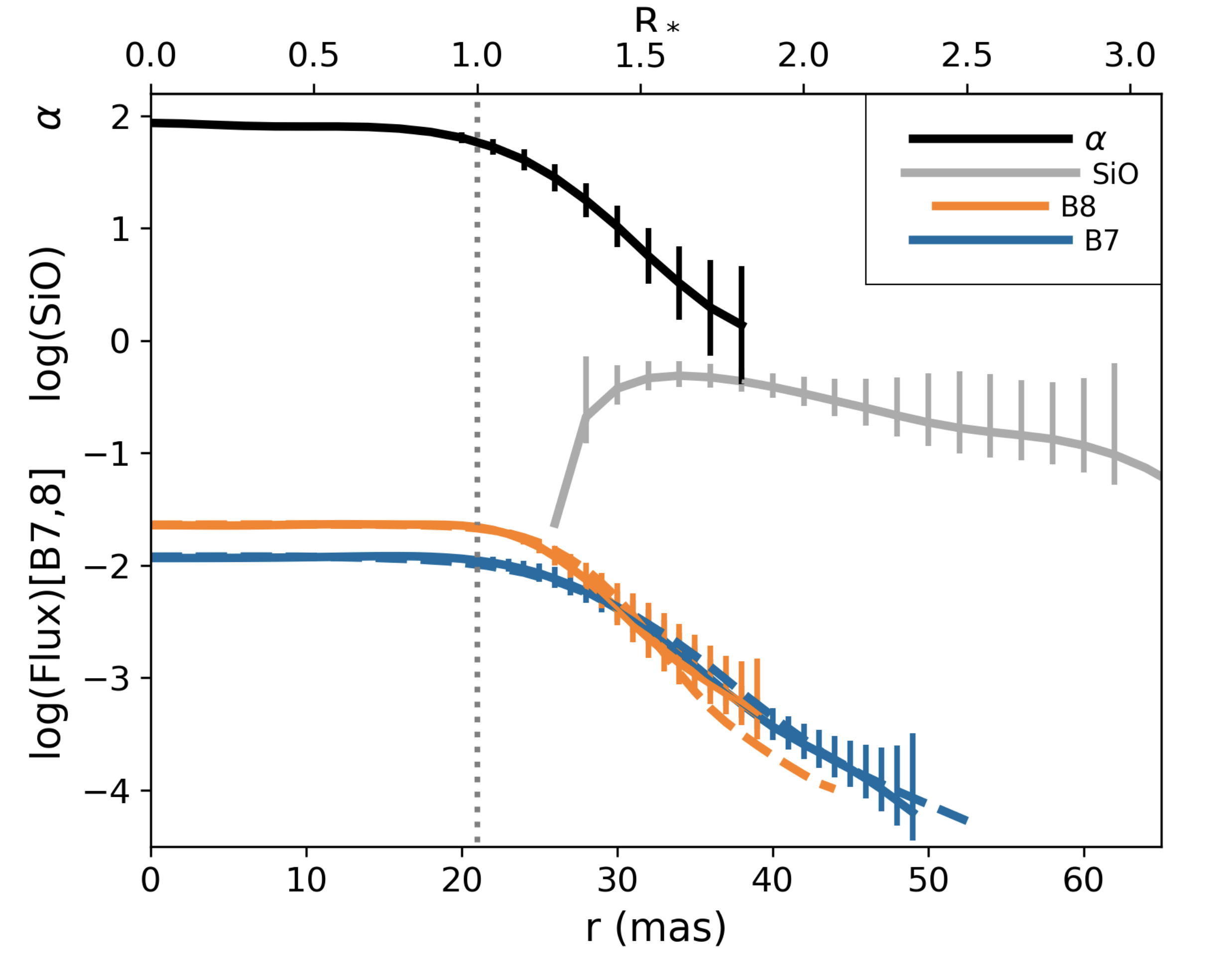}  
    \caption{Upper panel: Spectral index, $\alpha$, obtained from Band 7 and 8 images. Colour scale is from $\alpha=0-2.0$. Field of view is 110~mas, and outer regions where flux at either band is $\leq5\sigma$ are flagged out (in pale blue). Superimposed contours show Band 7 continuum (levels as in Fig.~\ref{fig:B7_cont_R0p5}), and the nominal 42~mas stellar diameter is shown by the dashed circle.
    Lower panel: Azimuthally-averaged radial distributions. Band 7 and 8 continuum fluxes are shown on a log scale by blue and orange solid lines, and corresponding dashed lines show the model with a simulated ALMA observation (Appendix~\ref{sec:appendix_sim}). Spectral index $\alpha$ is the upper black line (on a linear scale; Sect.~\ref{sec:cont_results}). Grey line is the distribution of integrated SiO emission ($v$=0 $J$=11-10) on a log scale; within 1.2~$R_{\star}$, SiO is in absorption (Sect.~\ref{sec:spec_lines}). Error bars include the noise and structural variations around each radial ring. Spatial resolutions are illustrated by bar lengths, upper right.}
	\label{fig:B78_spindex}
    \end{center}
\end{figure}

The Band 7 and 8 data were combined to derive the image of the spectral index $\alpha$, shown in Fig.~\ref{fig:B78_spindex} (upper panel). This was obtained from the ratio of images restored with the same circular 10~mas diameter beam, after correcting for the astrometric difference between the bands\footnote{Band 6 data were not used in this analysis because of the lower resolution and lack of {\it uv} overlap with the higher frequency data}. Two other methods for determining $\alpha$ were attempted: performing a dual-band \texttt{tclean} (selecting the same {\it uv} coverage in wavelengths) and using individual spectral windows within a single band. Both gave consistent results, although the S/N ratio using the latter in-band method was significantly lower due to the smaller frequency range.  
The azimuthally-averaged radial distribution of the fluxes and $\alpha$ are shown in the lower panel of Fig.~\ref{fig:B78_spindex}. Note that the error bars on $\alpha$ represent the relative uncertainty at each radius; they are considerably lower than the effect of the absolute flux uncertainty in the two bands,  which would result in a vertical shift to this curve.

Within the 42~mas diameter nominal stellar disk, $\alpha$ is relatively constant (shown by the dashed circle in Fig.~\ref{fig:B78_spindex}) and, applying the formal ALMA calibration uncertainties in Bands 7 and 8, we find $\alpha=1.97\pm$0.34. Using Bands 6 and 7, where the net calibration error is smaller, $\alpha=1.83\pm$0.21 in the same region. The combined value, $1.87\pm$0.18, is consistent with the blackbody limit of $\alpha=2$, indicating optically thick emission sampling the region of the atmosphere between the lukewarm chromosphere and the photosphere. This specific $\alpha$ is considerably higher than the value based on the total fluxes in a large aperture ($\sim$1.5; see above) as the source size is wavelength-dependent (see also \citealt{Matthews2026}).

As the emission in Bands 6 to 8 is expected to be characteristic of the respective $\tau_{\nu} \sim 1$ layers (from Eddington-Barbier relations), it is of interest to know
at which radii these layers correspond. The increase in axisymmetric angular size of the star towards longer wavelengths in Table~\ref{tab:uv_fits} can be used to estimate the mean $\tau_{\nu}=1$ radius with a simple analytical model for the specific intensities from a spherical turbulently-extended 
(${\rm v}_{turb}=14.0-19.5\>{\rm km\>s}^{-1}$, see Appendix~\ref{sec:appendix_model}) and quasi-static envelope. The density and opacity gradients are much smaller than the thermal gradient, so one can assume locally isothermal conditions. We adopt the stellar parameters for
Betelgeuse from \citet{Macleod2025}, and take the Rosseland $\tau=2/3$ angular diameter to be 1.03 \citep{Arroyo-Torres2013} times the 2.2$\mu$m continuum diameter of 42.61\,mas \citep{Montarges2021}, i.e. 43.89\,mas. 
Using the relationship between the free-free continuum optical depth in the line of sight towards disk-centre and that at the limb \citep{Menzel1936}, we find that the observed spectral index of $\sim$1.87 would refer to radii of $R\sim 1.16\to 1.20R_\star$, depending on the adopted turbulence velocity.

\begin{figure}
\begin{center}
     \includegraphics[width=7cm,angle=-90,trim=2cm 3cm 2cm 2cm,clip]
     {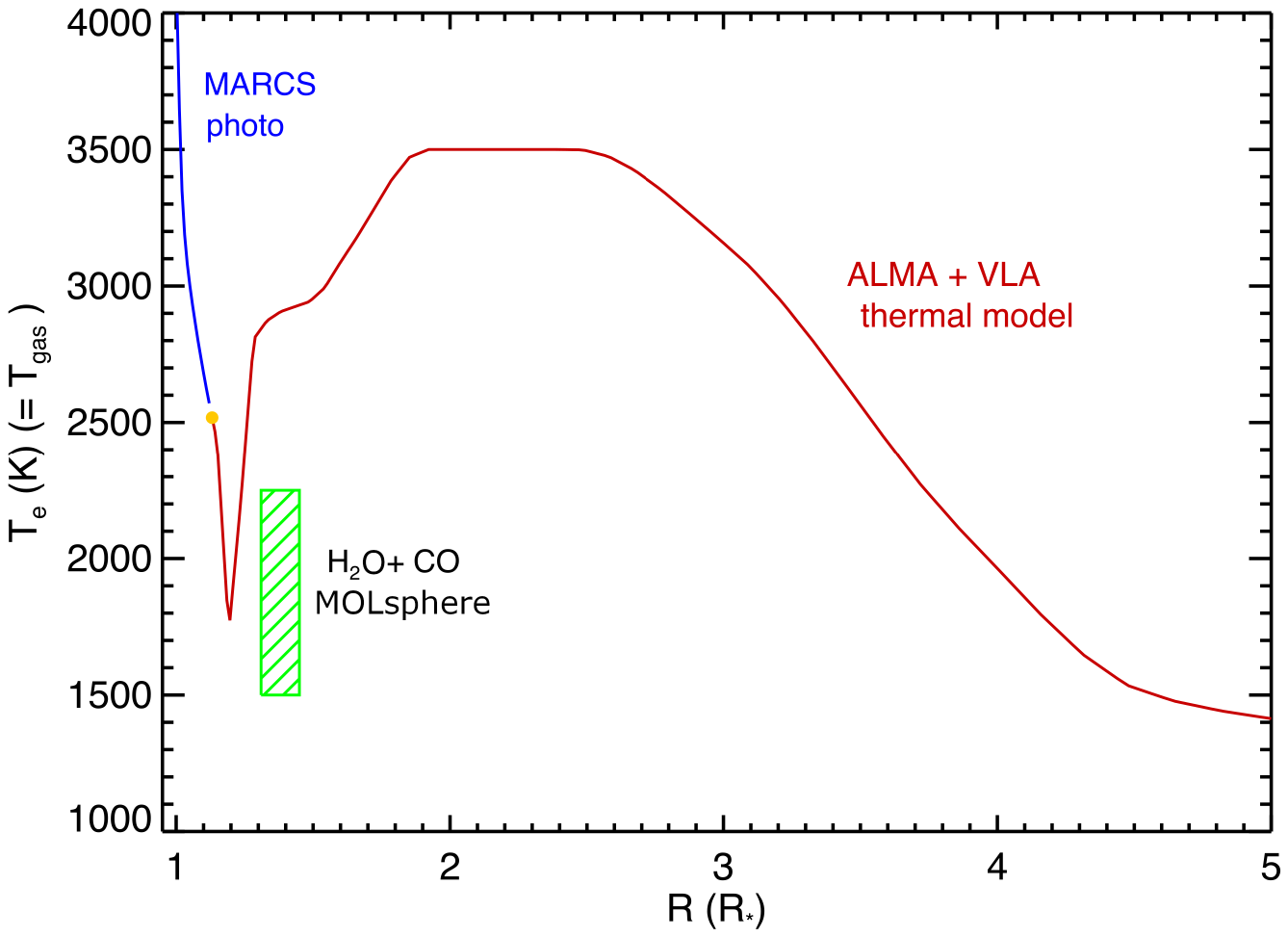}
    \caption{Thermal model (red) constructed from ALMA Bands 3, 4, 6, 7, and 8, and VLA Bands Q, K, U, X, C, and L data
as described in Appendix~\ref{sec:appendix_model}. The adopted MARCS photospheric model is shown in blue, and the join location is
the filled gold circle.  The ALMA Bands 6, 7, and 8 are formed in the temperature minimum region.
The lowest temperatures and link to the photospheric region are not well constrained because Betelgeuse
has not been been observed in ALMA Bands 9 and 10. Several studies have constructed simple models
for the molecular MOLsphere and its approximate location is given by the green shaded box; see Sect.~\ref{sec:radial_structure} and Appendix~\ref{sec:appendix_model} for details.
    }
	\label{fig:T_model}
    \end{center}
\end{figure}

A more rigorous approach for studying the formation of Band 6 through 8 emission is to construct a spherical
semi-empirical thermodynamic model (SEM) that accounts for all overlying chromospheric plasma; the emission cannot be independent of the free-free emission occurring at other sub-millimetre and centimetre wavelengths \citep{Harper2001}. We have constructed a
new SEM as described in Appendix~\ref{sec:appendix_model}, and the derived thermal structure shows a local temperature minimum around $\sim$1.2$R_\star$, before increasing again towards the stellar surface (Fig.~\ref{fig:T_model}; \citealt{OGorman2017}). It is the first to be
constructed using both ALMA and VLA visibility data, and this SEM was used to compute the model radial distributions shown as dashed lines in Fig.~\ref{fig:B78_spindex}.
From this model, the Band 7 and 8 spectral index corresponds to a disk-centre  $\tau=1$ radius of $1.14\>R_{\star}$.
Beyond the stellar limb, the observed $\alpha$ gradually decreases from 1.97 to $\sim$0.5 (Fig.~\ref{fig:B78_spindex}) as the optical depths decrease below unity, above the optically thin H$^-$ free-free limit of $\alpha=0.0$.  On the stellar sub-millimetre photosphere, the regions of enhanced $T_{\rm b}$ to the NE and SW show slightly lower $\alpha$ ($\sim$1.9) compared with the mid-plane ($\sim$2.0, see Fig.~\ref{fig:B78_spindex}). This may suggest slightly different vertical temperature gradients in these regions. 

Empirical radio brightness temperatures are effectively an average over a spatial volume, and near the temperature minimum
region they represent an upper limit to the actual minimum gas temperature in any theoretical model description.
In the Betelgeuse chromospheric model of \cite{Lobel2000} they find the temperature minimum to occur near 2769K for $R=1.1R_\star$ (for slightly different stellar parameters). \cite{Basri1981}
found a temperature minimum of 2730 K, from modelling of the Ca~II H \& K and Mg~II h \& k
line profiles, at a small radial extent of $<2$\% - but assumed a higher surface gravity. The
empirical ALMA brightness temperature values measured here, $\simeq 2300$\,K, are significantly lower and more consistent with surface-to-effective temperature ratios inferred from CO $4.6\>\mu$m Fundamental Band lines in K spectral-type red giant stars \citep{Wiedemann1994,Harper2013}, and VLA observations at 7~mm \citep{Matthews2022}.
The differences between the minimum temperatures of chromospheric models based on optical and ultraviolet diagnostics and those based on molecular lines have been interpreted as a thermal bifurcation between a molecular-cooled region where $T_{\rm min}\sim 0.6 T_{\rm eff}$ and a hotter mechanically-heated chromospheric region typically with a small filling factor. For Betelgeuse, this ratio would yield $T_{\rm min} \sim 2190$\,K, and such molecular cooling can be initiated by radiative instabilities in dynamical atmospheres \citep{Cuntz1994} and has also
been suggested to explain Betelgeuse's centimetre-radio fluctuations \citep{Drake1992}.

The first spatially-resolved radio multi-wavelength VLA chromospheric model \citep{Harper2001} also had a low $T_{\rm min}$ of 2400\,K at $R=1.08R_\ast$, where the $T_{\rm min}$ region was constrained with the then available spatially-unresolved published 1.2 and 3\,mm fluxes.
The new ALMA spatially resolved data now place much tighter constraints 
on the region between the chromosphere and the photosphere than previously possible and represent a significant improvement.
It appears that ALMA Bands 6, 7, and 8 are probing the cooler layers where molecular formation
will be enhanced. Interferometry has revealed the presence of CO, H$_2$O, and SiO molecules,
and also alumina in this region
(see, e.g. \citealt{Perrin2007}, \citealt{Ohnaka2009}, and \citealt{Gonzalez-Tora2024}, and Sect.~\ref{sec:spec_lines}).
The formation of CO and H$_2$O or alumina Al$_2$O$_3$ in these layers will not significantly alter the ALMA H$^-$ free-free opacities, but SiO may partially deplete the Si electron contribution.
At higher frequencies, ALMA Bands 9 and 10 may penetrate this temperature inversion zone, sampling deeper atmospheric layers. As the temperature will eventually start rising towards the photosphere, the spectral index may then gradually increase to $\gtrsim2$ \citep{Arbab2024}. The contribution of dust could also increase $\alpha$ above 2 at higher frequencies, although dust emission is thought to have a low optical depth and is thus not likely to be significant (see Appendix~\ref{sec:appendix_model}).

\subsubsection{Deviations from axisymmetry}
\label{radial_deviations}
\begin{figure}
\begin{center}
     \includegraphics[width=9.0cm,angle=0]
     {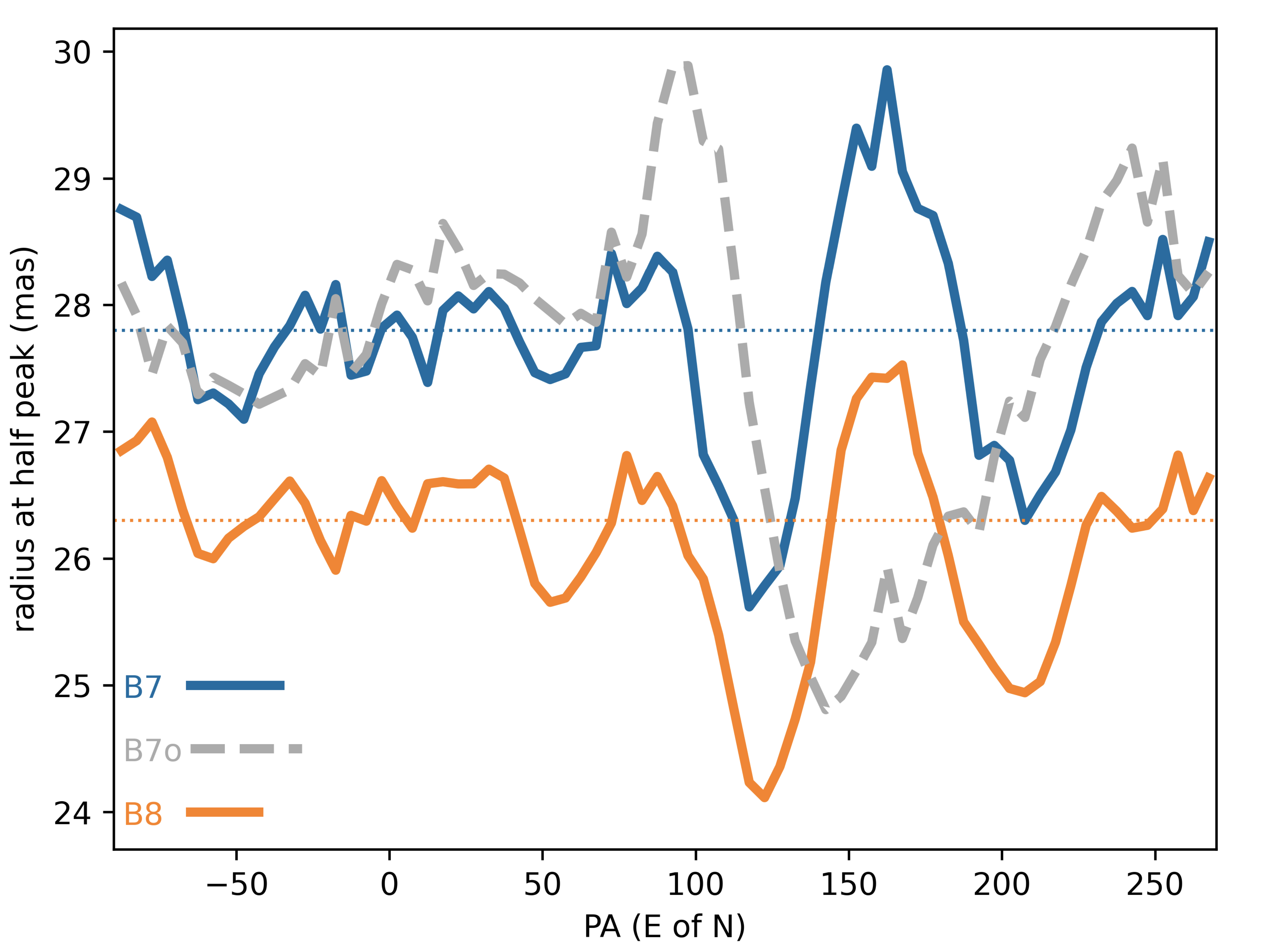}
    \caption{Radius of the continuum ($R_{0.5}$) in Bands 7 (blue) and 8 (orange) as a function of position angle. The radius from the emission centre (at half of the central flux) was fitted using cuts through images with 10$\times$10 and 7.7$\times$6.6mas restoring beams in the two bands. The effect on the derived radius from deconvolving with these elliptical beams is $\sim0.5$~mas at Band 7 and 0.2-0.3~mas at Band 8. Also shown by grey dashed line is $R_{0.5}$ at Band 7 measured in the 2015 dataset. Data were corrected for pointing offsets by centroiding the emission; remaining offsets would introduce a single-cycle modulation on these plots, and the good agreement between the bands in the 2023 data (apart from the expected fixed offset - see text) suggests these effects are small. The average spatial resolutions of the datasets are illustrated by the bar lengths, lower left.
    }
	\label{fig:radius_az}
    \end{center}
\end{figure}

At high resolution in all three bands, the bright central sub-millimetre photosphere appears non-circular (Fig.~\ref{fig:contin_images}). This is illustrated further in Fig.~\ref{fig:radius_az} where we show the radius, $R_{0.5}$, as a function of position angle for the highest resolution Bands 7 and 8 data. This was measured from cuts at different PAs, taking the radius at half of the median flux in the central 42~mas region. The variations are similar in the two bands, apart from a fixed vertical offset due to the higher optical depth at Band 7 (see above). The median values of $R_{0.5}$ are 27.8 and 26.3~mas at Bands 7 and 8, respectively, but the sector to the SSE (PA$\sim$160$^{\circ}$) has a radius 5\% larger, and to the SE (PA$\sim$120$^{\circ}$) it is 7\% smaller. The fractional change was found to be similar when measuring the radii at the 10\% level ($R_{0.1}$), and can be seen in the residual image in Fig.~\ref{fig:B7_cont_R0p5}, implying that these changes in the overall density or temperature extend beyond the immediate surface where $\tau\sim$1. 
Fig.\ref{fig:radius_az} also indicates that the scale of the radial corrugations is a significant fraction of the stellar circumference.

The grey dashed line in Fig.~\ref{fig:radius_az} is $R_{0.5}$ at Band 7 measured from the 2015 data \citep{OGorman2017}. It also shows ripples in the radius of the SSE sector, although the details of the structure appear to have changed.
In Sect.~\ref{sec:cont_discuss} we discuss possible origins of the surface structure in the continuum, and look at potential causes for the changes seen over the 7-year interval in Sect.~\ref{sec:cont_changes}.

\subsection{Spectral lines}
\label{sec:spec_lines}
As discussed in \cite{Kervella2018}, Betelgeuse shows many spectral lines of SiO and CO and their isotopologues, both in absorption against the millimetre and sub-millimetre continuum and in emission from the surrounding region.
Table~\ref{tab:line_results} identifies the lines detected in the present dataset, along with their integrated fluxes measured in absorption against the photosphere and in emission from a fixed annulus around the star. The integrated spectra are shown in Appendix~\ref{sec:appendix_spectra}. The frequencies given are as measured in the LSR frame by fitting the absorption line profile integrated over the stellar disk, and comparison of all the SiO and CO line frequencies (including isotopologues) with identified catalogue values from \textit{Splatalogue}\footnote{https://splatalogue.online}, gives a mean absorption velocity of ${\rm v}_{\mathrm{LSR}} = +2.4\pm$2.5~km s$^{-1}$. Compared with the published stellar velocity (${\rm v}_{\star}$~=+4.9~km~s$^{-1}$) this implies an average blue shift with respect to ${\rm v}_\star$ - consistent with a net gas outflow from the star \citep{Kervella2018}, but with an uncertainty of a few km s$^{-1}$ due to the irregular nature of the gas distribution and line shapes. The brighter SiO lines show more clearly that the absorption is blue-shifted (see position-velocity diagrams in Fig.~\ref{fig:pv_diagrams}, and discussion in Sect.~\ref{sec:discuss-velocities}; see also Appendix~\ref{sec:appendix_spectra} and \ref{sec:appendix_chanmap}). Apart from SiO and CO isotopologues, the only other lines detected were the Rydberg transitions in emission around 232~GHz, discussed further in \cite{Dent24}. The SiO lines with energy level E$_{low}$ up to 3500~K were detected in emission, whereas the lines up to 13800~K were seen in absorption. Note that the $v$=0 lines of CO are known to be extended over many arcsec (e.g. \citealt{OGorman2012}), and much of the emission is resolved out by current observations.  
The frequency of the $J$=1-0 line of [{\mbox{C\,{\sc i}}}] was included in the Band 8 data but the line was not detected, suggesting that the emission seen in a single-dish observation (14 arcsec beam, \citealt{vanderween1998}) comes from a much larger-scale structure which was resolved out in the ALMA observations.

\begin{table*}
	\centering
	\caption{Lines detected in absorption against photosphere, or emission in surrounding annulus. }
	\label{tab:line_results}
	\begin{tabular}{lccccllll}
		\hline
Line & Transition & E$_{low}$ & Meas. freq.$^{(1)}$ & Cat. freq.$^{(2)}$ & Emission$^{(3)}$ & Absorption$^{(4)}$ & rms$^{(5)}$ & Notes\\
 &  & (K) & (GHz) & (GHz) & (Jy.km~s$^{-1}$) & (Jy.km~s$^{-1}$) & (Jy.km~s$^{-1}$) & \\
\hline

SiO & v=2 5-4 & 3541 & 214.0899 & 214.0886 & 0.384 & 0.245 & 0.008 & Maser spot \\
$^{29}$SiO & v=0 5-4 & 20.6 & 214.3850 & 214.3857 & 0.604 & 0.454 & 0.007 & \\
SiO & v=1 5-4 & 1790 & 215.5970 & 215.5961 & 1.142 & 0.537 & 0.007 & Maser spikes \\
SiO & v=0 5-4 & 20.8 & 217.1046 & 217.1049 & 1.376 & 1.06 &  0.006&\\
$^{12}$CO & v=0 2-1 & 5.5 & 230.5368 & 230.538 & 0.644 & 0.261 & 0.006 & (7) \\
H30$\alpha$ & - & - & 231.9055 & 231.9009 & 0.572 &  -0.123$^{(6)}$ &  0.006 &Rydberg line$^{(6)}$ \\
X30$\alpha$ & - & - & 232.0255 & 232.0232 & 0.60 & -0.417$^{(6)}$ & 0.006 &Rydberg lines$^{(6)}$ \\
\hline
SiO & v=7 8-7 & 12082 & 330.4723 & 330.4775 & ($-$0.004) & 0.216 & 0.008 &(8) \\
$^{13}$CO & v=0 3-2 & 15.8 & 330.5831 & 330.5879 & 0.519 & 0.156 & 0.008 &(7,8)\\
$^{29}$SiO  & v=5 8-7 & 8677 & 331.1512 & 331.1581 & (0.125) & 0.086 & 0.008 & \\
$^{30}$SiO & v=3 8-7 & 5249 & 331.9500 & 331.9554 & (0.015) & 0.191 & 0.007 &(8) \\
SiO & v=6 8-7 & 10415 & 332.8742 & 332.8788 & ($-$0.020) & 0.597 & 0.008 & (8)\\
SiO & v=2 8-7 & 3579 & 342.5028 & 342.5044 & 1.173 & 1.66 & 0.006 & (8)\\
$^{12}$CO & v=1 3-2 & 3100 & 342.6408 & 342.6477 & ($-$0.079) & 0.14 & 0.006 &(8)\\
$^{29}$SiO & v=0 8-7 & 57.6 & 342.9784 & 342.9791 & 2.670 & 1.57 & 0.006 &(8) \\
SiO & v=1 8-7 & 1827 & 344.9149 & 344.9163 & 2.867 & 2.23 & 0.007 &(8)\\
$^{12}$CO & v=0 3-2 & 16.6 & 345.7935 & 345.7960 & 2.657 & 0.925 & 0.008 & (7,8)\\
\hline
SiO & v=0 11-10 & 115 & 477.5043 & 477.5052 & 9.30 & 8.44 & 0.04 &\\
{[\mbox{C\,{\sc i}}]} & 1-0 & - & - & 492.1607 & (0.74) & (0.074) & 0.04 & Not detected$^{(9)}$ \\
SiO & v=8 12-11 & 13806 & 491.9941 & 492.0021 & (0.68) & 1.83 & 0.04 & \\

\hline

\end{tabular}

(1) Measured centre frequency of Gaussian fit to absorption line, with errors of typically 0.1-0.2~MHz for the brightest transitions, up to $\sim$0.5~MHz for the faintest; 
(2) Catalogue frequencies are from the CDMS or JPL databases \citep{Pickett1998,Muller2001}, as reported by \textit{Splatalogue}$^8$;
(3) Integrated line emission over $\pm$20~kms$^{-1}$ centred on target velocity using line rest frequency, measured in an annulus of radii 44 to 88~mas (non-detections given in brackets); 
(4) Integrated line absorption over $\pm$20~kms$^{-1}$ measured in a 22~mas radius aperture centred on the continuum position;
(5) Noise measured in 40~kms$^{-1}$ channel bins, not including additional uncertainty due to calibration and spectral baseline (continuum subtraction) errors;
(6) Rydberg lines of hydrogen and heavier elements are in emission rather than absorption against the star, discussed in \cite{Dent24};
(7) Emission fluxes are a lower limit, as much of the extended gas is resolved out \citep{OGorman2012};
(8) Also observed in 2015 \citep{Kervella2018};
(9) Integrated values are over the same range as the detected lines, centred on the catalogue frequency.

\end{table*}

\begin{figure}
\begin{center}
     \includegraphics[width=7.1cm,angle=0]
     {Fig7a_SiOB7.pdf}
     \includegraphics[width=7.1cm,angle=0]
     {Fig7b_COB7.pdf}
     \includegraphics[width=7.1cm,angle=0]
     {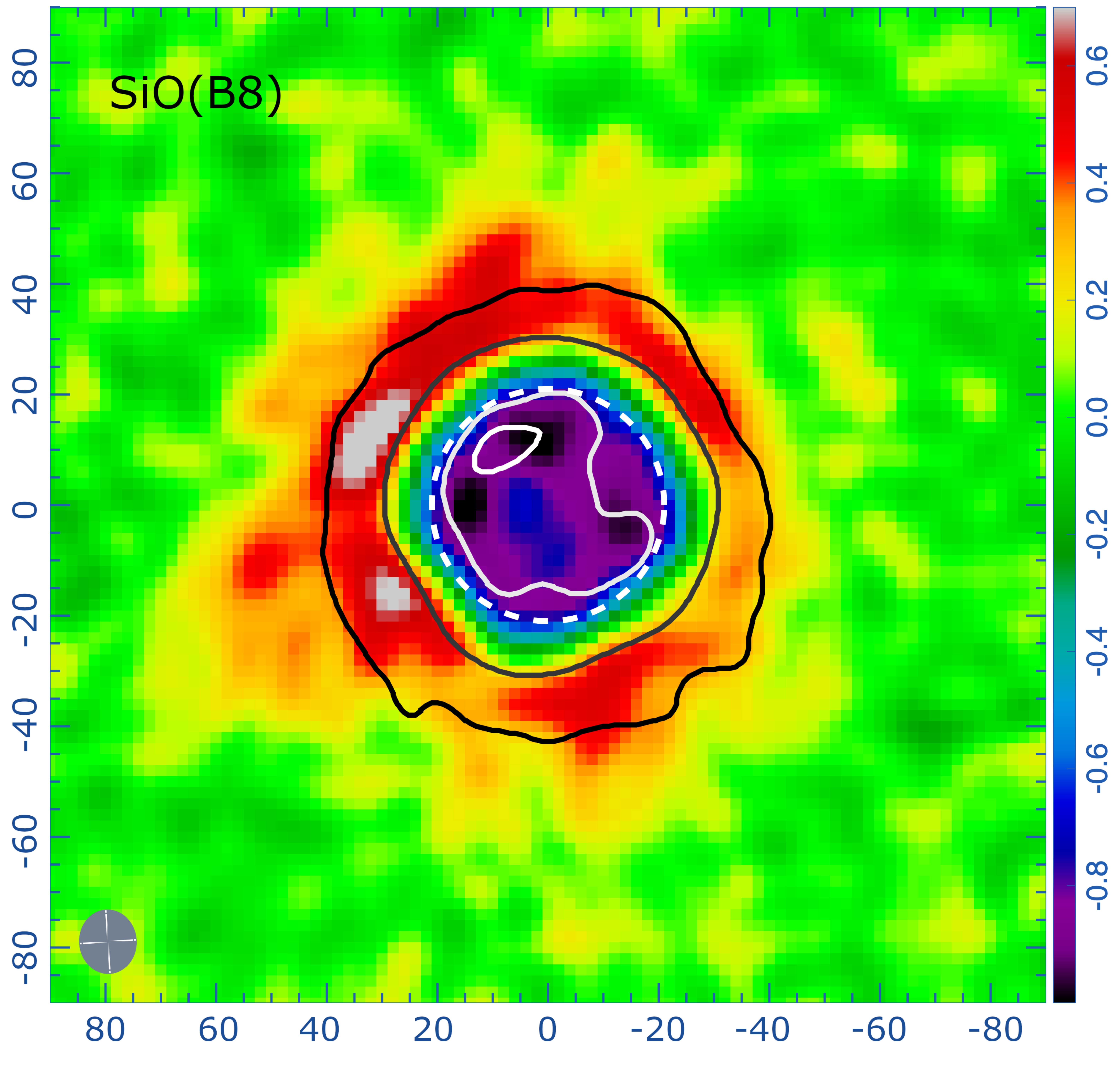}
    \caption{Integrated line emission and absorption, in SiO $v$=1 $J$=8-7 (B7), CO $v$=0 $J$=3-2 (B7), and SiO $v$=0 $J$=11-10 (B8). Line absorption against the continuum is shown by the blue through black colour table, and emission is shown as yellow through white. Velocity range is $\pm$30~km s$^{-1}$. Superimposed solid contours show continuum in the respective bands (B7 levels as in Fig.~\ref{fig:B7_cont_R0p5}; B8: 0.8,10,40,44~mJy/beam). Dashed contours in B7 images are line emission starting at 3$\sigma$ (levels +0.05,0.07,0.25,0.3~Jy/beam km s$^{-1}$). Field of view is 180~mas, beam sizes are lower left, and the 42~mas stellar diameter is shown by the dashed white circle. 
    }
	\label{fig:integ_SiO}
    \end{center}
\end{figure}

The spatial and velocity distributions of the spectral lines appear quite similar in the $J$=8-7 and 11-10 transitions of SiO and its isotopologues, and in CO. This is illustrated in Fig.~\ref{fig:integ_SiO} and in the channel maps in Figs.~\ref{fig:SiO_B7_channel}--\ref{fig:CO_channel}. Higher resolutions reduce beam confusion at the stellar limb between line absorption against the continuum and emission from the surrounding region, so the Band 8 emission peaks around 33~mas radius (1.6~$R_\star$), compared with 37.8~mas in Band 7 \citep{Kervella2018}. The azimuthally-averaged Band 8 SiO radial profile is shown in grey in Fig.~\ref{fig:B78_spindex}.

The $J$=5-4 $^{28}$SiO $v$=1 and $v$=2 emission in Band 6 are dominated by a compact peak around ${\rm v}_{LSR}$=10.5—11.5 km s$^{-1}$ (see channel maps, Fig.~\ref{fig:SiO_B6_channel}).  This lies about 33~mas from the centre of the star at position angle $\approx60\pm10^{\circ}$, similar to the hot-spot and axis direction.
The maximum $J$=5-4 flux densities at 11.5 km$^{-1}$ for $v$=0, 1, 2 at the position of the peak are 0.0104, 0.153, 0.0185 Jy beam$^{-1}$, corresponding to absolute brightness temperatures $T_{\rm b}$, of 400, 5860, 710 K $\pm 40$ K (lower limits since the maser spots may be much smaller than the beam).   Assuming optically thin emission and a Boltzmann level population distribution, then the ratio of line fluxes in the same J transition from the same molecule is expected to
be given by $A_{\mathrm{UL}i} / A_{\mathrm{UL}j} \times exp((E_{\mathrm{u}i} - E_{\mathrm{u}j}) / T)$ for vibrational states $i, j$ at temperature $T$, taken as 2000 K. $A_{\mathrm{UL}}$ and $E_{\mathrm{u}}$ are the Einstein A coefficient and upper energy level, in temperature units.  This predicts the ratios  for $v$=1 and $v$=2 relative to $v$=0 of 0.413 and 0.172.  The measured $T_{\rm b}$ ratios are 12.5 and 1.8 (uncertainty $\sim0.2$).  Thus, both vibrationally excited $J$=5-4 lines are significantly brighter than predicted and are likely to be masing. The compact size and narrow spectral profile of the $v$=1 emission are consistent with this, although a higher resolution would be needed to confirm this for both lines.  

The distributions of the $J$=8-7 $^{28}$SiO $v$=1 and $v$=2 emission are more extended and are quite similar to those of CO (Figs.~\ref{fig:integ_SiO}, \ref{fig:SiO_B7_channel} and \ref{fig:CO_channel}). The $v$=0 was not observed so the same comparison cannot be made, although the brightness ratios measured in 2015 suggested that $v$=1, 2 were masing \citep{Kervella2018}. Comparing the present integrated Band 7 SiO emission map in Fig.~\ref{fig:integ_SiO} with the image of the same line from 2015 (Fig.5 in \citealt{Kervella2018}) shows that the distribution of the clumps has changed over the intervening 7 years. The exceptionally bright maser spot observed in the new $J$=5-4 data in $v$=1, 2 is not surprising as the levels and thus pumping requirements involved in masing are different for the different transitions.   
There is no evidence of masing of the other SiO isotopologues observed here, and the $v$>2 states are detected in absorption.

As CO is unlikely to be masing, we assume that the emission of both CO and the SiO $J$=8-7 transitions that follow it in the 2023 dataset trace the MOLsphere above the optically thick sub-millimetre photosphere \citep{Kervella2018}. Although the radial extent of free-free and line emission is similar, Fig.~\ref{fig:integ_SiO} shows that the molecular gas is significantly more clumpy, with a NE-SW contrast as high as $\sim$10:1 at a radius of 30~mas, compared with variations of only $\sim$10\% in the continuum (Fig.~\ref{fig:B7_cont_R0p5}). A possible origin of this weak emission in the SW sector is discussed in Sect.~\ref{sec:cont_changes}. Channel maps (Appendix~\ref{sec:appendix_chanmap}) and position-velocity (PV) cuts (Fig.~\ref{fig:pv_diagrams}) also show that these clumps can have different projected velocities.

Molecular line absorption against the stellar continuum also appears clumpy (shown by the region within the dashed circle in Fig.~\ref{fig:integ_SiO}), and is stronger towards the stellar limb. This might be expected because of the longer line of sight due to the geometry of a radially-extended turbulent absorbing layer. The hotter continuum regions also appear to have the deepest line absorption, and this possible association is discussed in Sect.~\ref{sec:cont_discuss}. The deepest absorption lines against the star (Appendix~\ref{sec:appendix_spectra}) have minimum temperatures $T_{\rm b} = -$(800--1200)~K in Bands 7 and 8, which implies a minimum apparent brightness temperature of $\sim$1300~K after correcting for the $\sim$2300~K continuum. For optically thick lines from molecular gas in local thermodynamic equilibrium (LTE) with a unity filling factor, this would equal the gas kinetic temperature, $T_{\rm g}$. However, if either the optical depth or filling factor was less than 1.0, the mean $T_{\rm b}$ against the star would increase, which implies that 1300~K is an upper limit to $T_{\rm g}$. This is comparable to or even lower than the estimated MOLsphere temperature in Fig.~\ref{fig:T_model}, suggesting that the absorbing molecular gas has both a high filling factor and an optical depth $\gtrsim$1.0.

\section{Discussion}

\subsection{Stellar disk structure}
\label{sec:cont_discuss}

The hot regions on the stellar disk and the variations of radius are seen in Betelgeuse at all three millimetre and sub-millimetre bands. The $\tau \sim 1$ surface at each wavelength samples a different radius (Table \ref{tab:uv_fits}), which implies that the hot patches and structural differences extend over a radial range of at least 4~mas or 0.2$R_\star$ through the inner atmosphere (or sub-millimetre photosphere). 
The locations of the two hot regions in Fig.~\ref{fig:contin_images} are similar to those derived through a visibility model of near-infrared observations taken in 2008 \citep{Haubois2009} (their Figure 10), suggesting that these features may extend all the way down to the optical/infrared photosphere. Optical observations with 14~mas resolution using VLT/SPHERE in 2018-2020 also suggested enhanced emission in the NE sector of the stellar disk \citep{Montarges2021}. However, optical spectropolarimetry suggested a different distribution of surface spots \citep{Auriere2016}, and the link between optical/infrared patches and those seen in the sub-millimetre is not yet proven.

\cite{Kervella2018} suggested that the higher temperature in the NE patch is due to a greater scale height of emission in that region, as the gas temperature $T_{\rm g}$ is thought to increase with height up to 2$R_\star$ (Fig.~\ref{fig:T_model}). The peak brightness temperature measured in the new data in all three bands reaches 3000~K, approaching that of the stellar surface and the derived $T_{\rm g}$ at 2$R_\star$. 
 
The RSG stars such as Betelgeuse are predicted to be dominated by a small number ($\sim$3-5) of large-scale convective cells \citep{Schwarzschild1975, Freytag2002, Chiavassa2024}. The release of energy from convection at the stellar surface is proposed to drive shocks into the atmosphere and is a suggested mechanism to enhance the stellar wind and mass loss from these stars \citep{Josselin_2007, Humphreys2022}. From the angular size and temperature excess of the NE hot patch, we estimate its luminosity to be $\sim$1\% of the star at these wavelengths, similar to the fractional energy carried to the surface in models of rising gas blobs \citep{Freytag2024}. Large convective cells can also explain some of the stellar variability \citep{Stothers2010}, the spotty and uneven surface of RSGs \citep{Haubois2009}, and the variable line shapes and polarisation \citep{LopezAriste2018}. If the NE and SW regions are also manifestations of large convective cells projected into the atmosphere, we can look at other aspects of the ALMA data in these regions, in particular spectral lines in the nearby MOLsphere and the lifetimes of the hot patches.

\begin{figure}
\begin{center}
     \includegraphics[width=10cm,angle=-90,trim=2cm 4cm 0cm 5cm,clip]
     {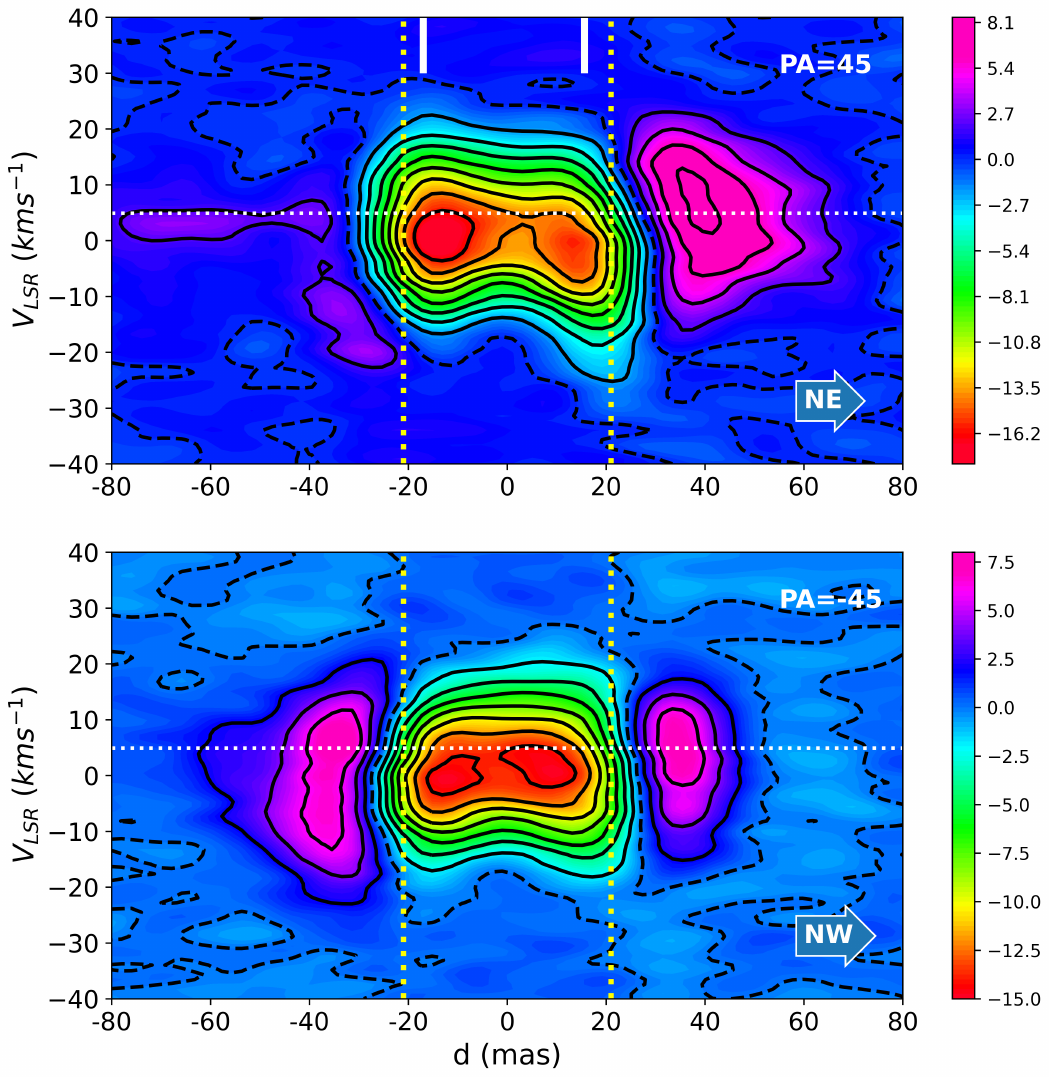}
    \caption{PV diagrams of the three brightest SiO lines in Band 7 combined (SiO $v$=1 $J$=8-7, SiO $v$=2 $J$=8-7, and $^{29}$SiO $v$=0 $J$=8-7), with cuts taken at PA=45$^{\circ}$ approximately through the two continuum hot patches (upper panel) and orthogonal to this direction (lower panel). All three lines have similar intensities and distributions, and were averaged together to increase the S/N. Vertical scale is velocity in the LSR frame. Dotted vertical lines delineate the nominal stellar radius (21~mas), solid vertical lines in upper panel show the offsets of the two continuum peaks, and horizontal dotted line indicates ${\rm v}_\star$. Green through red colour scale represents absorption against the photosphere, and pink is emission, with the scale in mJy/beam. Contours steps are 2mJy/beam (=5$\sigma$), with zero shown by the dashed contour. The cuts were averaged over 3 pixels (6~mas).
    }
	\label{fig:pv_diagrams}
    \end{center}
\end{figure}

The integrated molecular line absorption is stronger near the bright continuum patches (Fig.~\ref{fig:integ_SiO}), with the contrast of the integrated SiO absorption in the NE compared to the centre reaching $\sim$40\%, significantly larger than the continuum contrast ($\sim$10\%). Additionally, the linewidths are broader near these regions (indicated by the vertical bars in the upper position-velocity diagram in Fig.~\ref{fig:pv_diagrams}). The free-free continuum emission forms an optically thick warm background screen at 1.15$R_\star$ (Sect.~\ref{sec:surface_structure}), with most molecular absorption in the MOLsphere at $\sim$1.3$R_\star$. Convective cell overshoot in the stellar surface would propagate shocks into the atmosphere and the MOLsphere at $5-7\times$ the sound speed ($c_s\sim$5 km s$^{-1}$, \citealt{Freytag2002}), which would be seen as blue-shifted absorption against the star. \cite{LopezAriste2018} suggested that the brightness in these cells should then be correlated with an increased blue-shifted gas, and that most of their detected absorption lines are from such upwelling gas. Lines in emission, however, should be equally distributed around the star and observed at separations >$R_{\star}$, and consequently average to the mean stellar velocity (${\rm v}_\star$ = 4.9~km~s$^{-1}$, \citealt{Kervella2018}). The linewidths in the present data are broad over the whole disk (10-20 km s$^{-1}$) and, as noted by \cite{Kervella2018} and \cite{Pilate2024} (also $\mu$Cep, \citealt{LopezAriste2023}), such broad red- and blue-shifted absorption in the face-on disk centre is surprising, as it would indicate no net up or down gas motion. Molecular lines in absorption against Betelgeuse (Fig.~\ref{fig:pv_diagrams}) do show a bias towards the blue-shift with respect to the star. Moreover, the location of the NE hot patch (indicated by the vertical dashed line at +14~mas in the upper panel) is close to the highest velocity blue-shifted gas, with absorption extending to $\sim-30$km s$^{-1}$ relative to ${\rm v}_\star$.  Hydrodynamic 3D models have suggested that surface velocities are more complex than simple radial motion \citep{Goldberg2022}, with both infall and outflow occurring at different times, and variations in the temperature structure as a function of optical depth. Thus, \cite{Freytag2024} concluded that hot regions may not always be associated with convective outflow. Deeper spectra at high spatial resolution at future epochs would be necessary to confirm whether the hotter continuum patches are consistently associated with blue-shifted upwelling gas.

The non-symmetrical shape of the half-power continuum contours (Fig.~\ref{fig:radius_az}) shows that the inner atmosphere is not spherical, with radial corrugations ($\delta R/R_\star$) of $\pm$6\% confined to the SSE sector. This is similar to the fractional size of convective plumes from optical observations and models, which propose radial fluctuations of 0.1$R_\star$ \citep{LopezAriste2023, Chiavassa2024}, and comparable to the radial variations seen in the atmosphere of the asymptotic giant branch (AGB) star R~Dor at millimetre wavelengths \citep{Vlemmings2024}. If the mean radial run of temperature in Fig.~\ref{fig:T_model} applied throughout, the sharp temperature inversion around 1.2~$R_\star$ along with a 6\% variation in radius could account for the hot patches on the stellar disk. The subtle change in the spectral index near these regions may also be indicative of a localised change in the vertical temperature structure.

\subsection{Temporal changes in structure}
\label{sec:cont_changes}
\begin{figure}[ht!]

\begin{center}
	\includegraphics[width=7.1cm,angle=0]{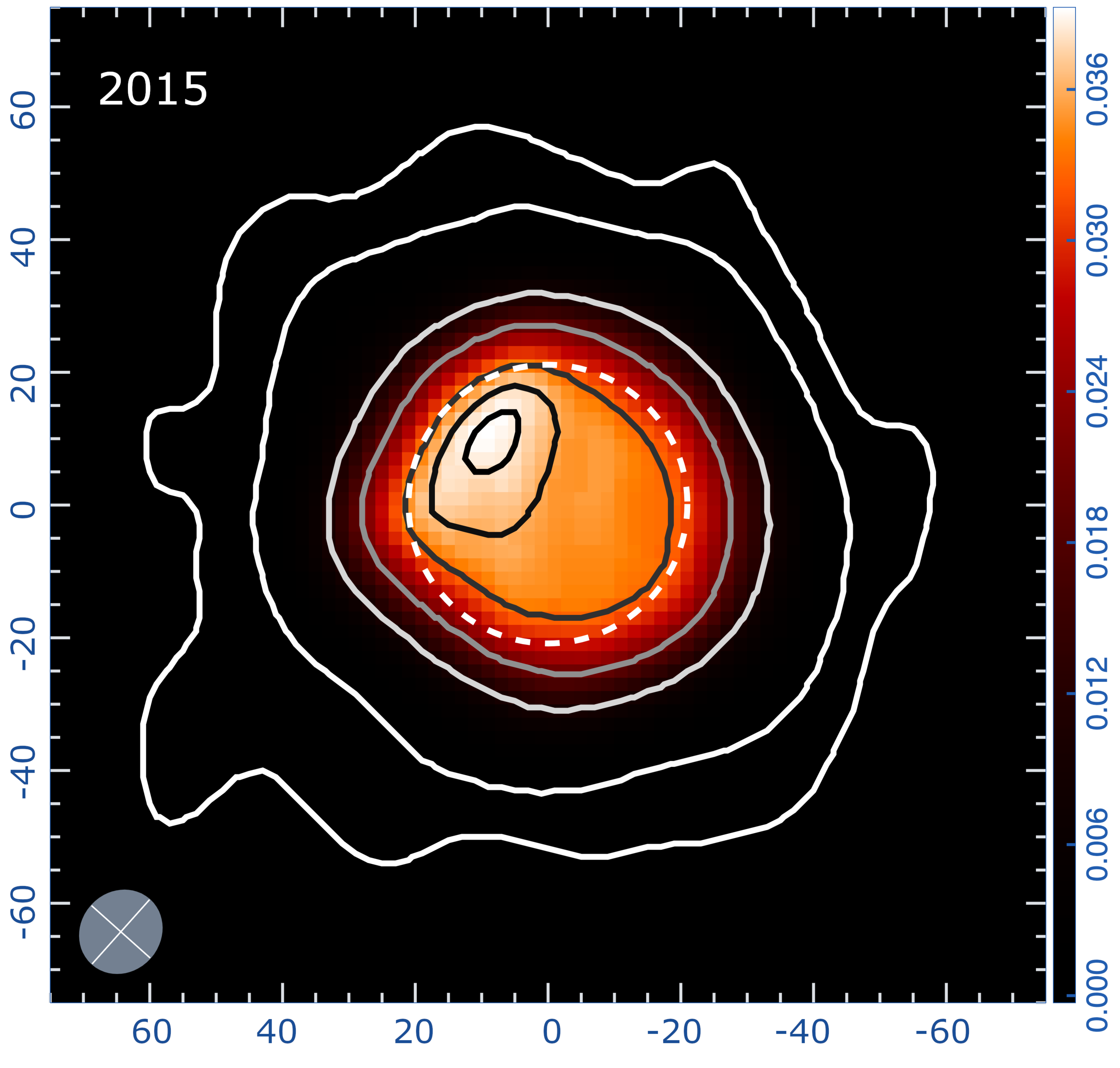}
    \includegraphics[width=7.1cm,angle=0]{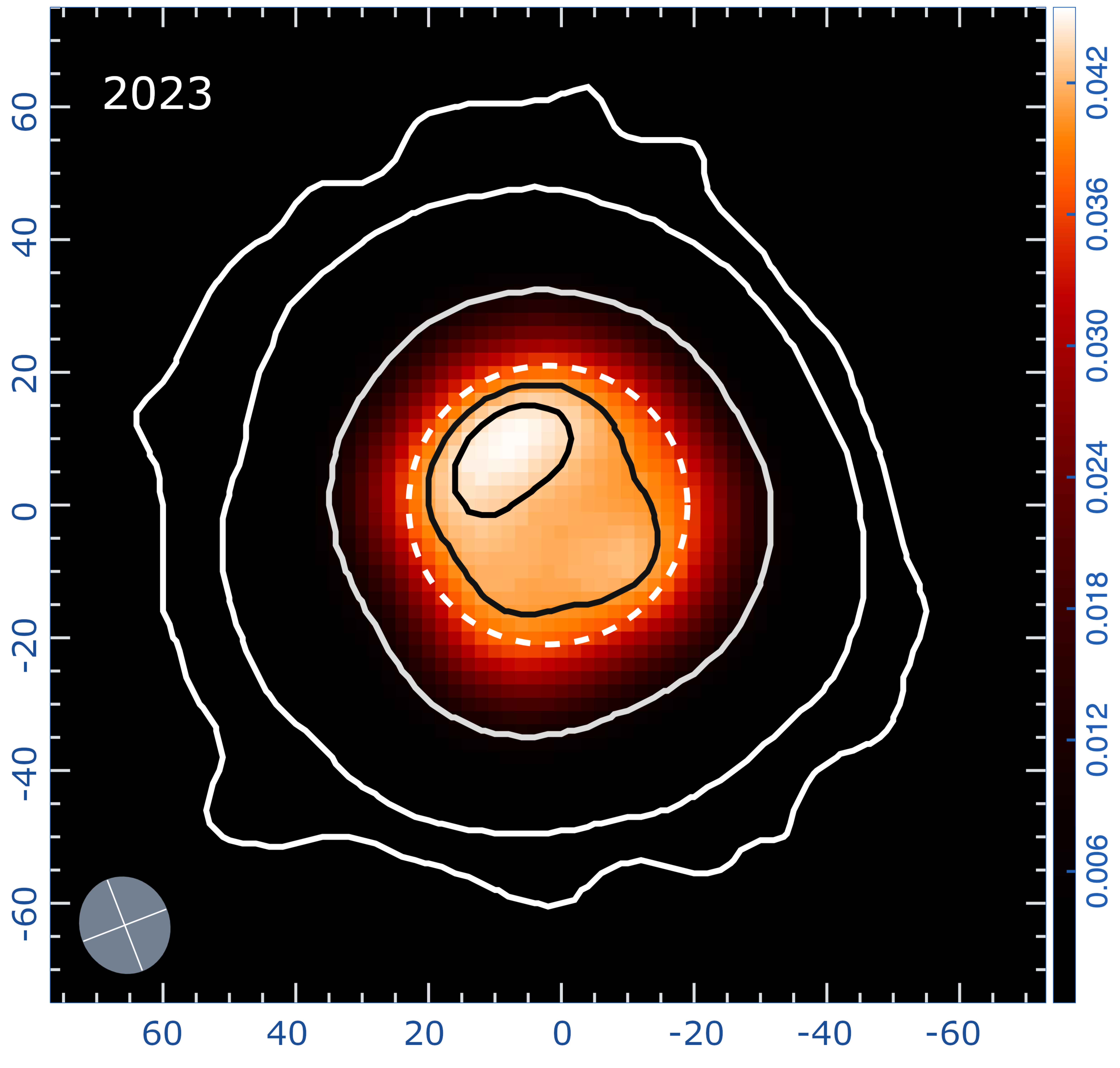}
    \includegraphics[width=7.1cm,angle=0]{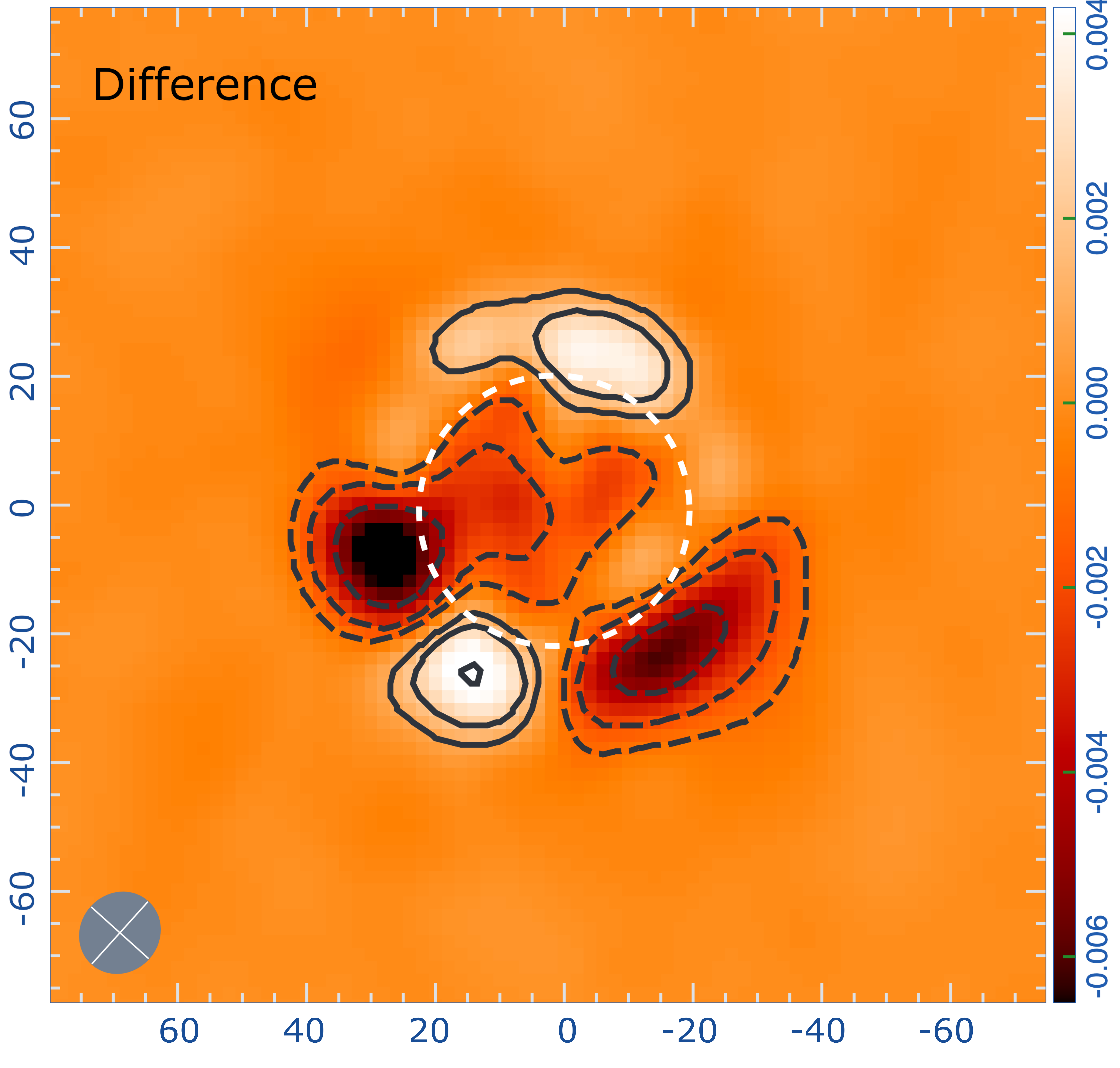}
     \caption{Comparison of Band 7 continuum observed in November 2015 \citep{OGorman2017} with that of August 2023. Lower panel is the difference image (2023-2015). Dashed circles represent the 42~mas stellar diameter. Colour scales are Jy/beam. The beam dimensions are shown lower left. Contour levels (starting at 5$\sigma$), are 0.25,1,10,20,30,36,38~mJy/beam in the upper panel, same as Fig.~\ref{fig:B7_cont_R0p5} in the centre, and $\pm$4,$\pm$2,$\pm$1~mJy/beam in the lower panel.
    }
	\label{fig:B7_compared}
    \end{center}
\end{figure}

The lifetimes of the structures on and around Betelgeuse may provide further clues as to their possible origin. ALMA Band 7 observations taken in November 2015 \citep{OGorman2017} used a long baseline configuration similar to the present dataset. Fig.~\ref{fig:B7_compared} compares these continuum data (re-imaged using super-uniform weighting) with the present image. Both show a hot NE region, and $\it uv$-fitting shows that the contrast and offset from the stellar centre have not changed significantly over the 7.3~year interval: \cite{OGorman2017} measured an offset from centre in Band 7 of (9.4, 12.9)~mas for the NE peak, within 2~mas of the present data (Table~\ref{tab:uv_fits}). The difference image (lower panel) shows that the most significant changes are at the limbs of the emission, which can also be seen in the radius versus azimuth plots in Fig.~\ref{fig:radius_az}.

The similarity of the sub-millimetre peak in 2015 and 2023 suggests that its lifespan is at least 7 years. 
\cite{Haubois2009} found two near-infrared hotspots with similar locations. The optical images in \cite{Montarges2021} suggest enhanced emission on the NE side of the disk. \cite{Gilliland1996} and \cite{Uitenbroek1998} also found a UV hotspot to the SW. If these are indeed tracing the same hot regions as seen at millimeter and sub-millimetre wavelengths, it would imply that these structures are even longer-lived ($\sim$20~years) and extend from the optical photosphere through the inner atmosphere of the star.
This timescale is somewhat longer than estimates from models of turbulent convection, which tend to show coherent large-scale lifetimes of $\sim$months to a few years \citep{Freytag2002, Ma2024, Freytag2024}. Features seen at millimetre wavelengths on the AGB star R~Dor also appear to have a comparatively short timescale, of months \citep{Vlemmings2024}.

The location of the dominant hotspot, measured as PA = 51$\pm$2$^{\circ}$ in the present data, was suggested by \cite{Kervella2018} to trace the orientation of the stellar rotation axis.  In which case, the NE hot patch may be a persistent convective feature close to the stellar pole. If it lies at the pole, its projected position on the disk would suggest a polar axis inclined towards us at $\sim30^{\circ}$ to the plane of the sky. However, an inclined polar origin does not readily account for the second hot region to the SW, unless this is the limb of an extended region on the far side. 

Unlike the hot patches on the disk face, the radial corrugations as a function of azimuth have changed over the 7-year period, although they are mostly confined to the same SSE sector (Fig.~\ref{fig:radius_az}). If driven by convective plumes, it suggests that the corrugations have a shorter lifetime. The longevity of these two types of feature should be examined in further epochs at high resolution.

The ALMA datasets taken in 2015 and 2023 bracket the 'Great Dimming' event, a 1.6~magnitude decrease in visual brightness lasting a few weeks in 2020. Optical images taken during the event \citep{Montarges2021} suggested that a cool spot and/or a dusty clump temporarily obscured the SW of the star, possibly associated with enhanced mass loss. Its orientation is similar to the weakest sector of the $\sim$30~mas radius MOLsphere (at PA $\sim$250$^{\circ}$ in Fig.~\ref{fig:integ_SiO}).
This emission gap is less clear in the 2015 integrated maps \citep{Kervella2018}, and although we only have observations at two epochs, it is possible that the 2020 event may have suppressed molecular formation and/or line emission in that sector of the MOLsphere. Additionally, at a radius of $\sim$60~mas at PA $\sim$220$^{\circ}$, a separate clump of weak molecular emission can be seen in Band 7 (Fig.~\ref{fig:integ_SiO}). The PV cut in Fig.~\ref{fig:pv_diagrams} (upper panel, at offset -60~mas) shows that this gas has a relatively narrow linewidth, with v$_{LSR}\sim 3\pm 3$~km s$^{-1}$. If this is tracing gas ejected during the Great Dimming, it would support the model of enhanced mass loss in a dusty clump. Assuming motion mostly in the plane of the sky - as supported by its low velocity relative to ${\rm v}_\star$ - the separation from the MOLsphere radius in the present data would imply a mean outflow velocity of $\sim$10~km s$^{-1}$.

Near-infrared and optical observations over the past $\sim$40~years show emission extending up to $\sim$120~mas to the SW of Betelgeuse (discussed in \citealt{Kervella2009}). This suggested a preferred direction for mass loss. The similarity to the orientation of the hot continuum patches on the sub-millimetre photosphere, the gap in the MOLsphere, and the separated SW molecular clump suggests a link between these phenomena. Further observations at a later epoch would be of interest in looking for proper motion of the separated clump, as well as the possible evolution of the MOLsphere.

\subsection{Gas velocities}
\label{sec:discuss-velocities}
The molecular lines in the vicinity of the star are dominated by a combination of absorption observed against the sub-millimetre photosphere and clumpy emission around the star at up to $\sim$3$R_\star$. The velocity structure is complex, and some contribution has been ascribed to rotation around an axis inclined approximately NE-SW, using sub-millimetre wavelength data with models to compensate for the effects of absorption \citep{Kervella2018} and at UV wavelengths on a larger scale \citep{Uitenbroek1998}. The position-velocity diagrams in Fig.~\ref{fig:pv_diagrams} show SiO lines in Band 7 taken at an angle approximately along the proposed rotation axis (upper panel, at an angle linking the two hot patches) and orthogonal to this (lower panel, where the rotation velocity should be most obvious). As all the observed Band 7 SiO lines have similar characteristic structures, the three brightest lines were combined to increase the S/N in these figures. There is no clear SE-NW velocity gradient in emission or absorption in the present dataset (lower panel in Fig.~\ref{fig:pv_diagrams}) as would be expected from rotation; if anything, the SE emission is more blue-shifted - opposite to that found in 2015. The upper panel (at the orthogonal angle, PA=45$^{\circ}$) does show an apparent velocity gradient in gas emission on the two sides of the star. However, the differences compared with 2015 suggest that there have been significant changes in the gas plumes (Sect.~\ref{sec:spec_lines}). This confirms that the MOLsphere gas is a complex structure that varies on timescales of $\sim$yr, possibly faster than variations in the continuum, and further analysis and observations would be necessary to investigate signatures of rotation.

\subsection{A binary companion?}
\label{sec:discuss-companion}

Betelgeuse has been suggested to have a companion, proposed to explain the 5.78~yr periodicity amongst other phenomena \citep{Macleod2025}. A candidate identified in coronographic observations by \cite{Howell2025} was recently confirmed at the 5-6$\sigma$ level by \cite{Montarges2026}. Based on available data, estimates of orbital elements \citep{Macleod2025} suggest that the companion's orbit is viewed approximately edge-on, with a projected orbital axis at PA$\sim$60$^{\circ}$ and a semi-major axis of 2.3~$R_\star$. \cite{Macleod2025} investigated possible effects of such a companion on the primary star, in particular the local excess luminosity and tidal disruption. \cite{Dupree2026} also found evidence of a tidal tail in the periodicity of the absorption lines.
The date of the present ALMA Band 7 observations (2023.59) was 0.47$\pm$0.34~yr after conjunction (the error being due to uncertainty in the orbital elements), giving an apparent separation of 1.1$\pm$0.7~$R_\star$ from the stellar centre at PA$\sim$150$^{\circ}$. The mean proposed location is shown by the black dot in the continuum in Fig.~\ref{fig:B7_cont_R0p5}, which places it too close to the limb to be detectable in the present data.

The angle of the proposed orbital axis of the companion (60$^{\circ}$) lies close to the angle between the NE and SW hot patches (51$^{\circ}$, see Fig.~\ref{fig:contin_images}), and orthogonal to the direction of elongation of the NE hot region (Table~\ref{tab:uv_fits}) and the enhancement in $\alpha$ (Fig.~\ref{fig:B78_spindex}). At a PA of $\sim$150$^{\circ}$, the 2023 location of the companion was close to where the sub-millimetre photosphere appeared most distorted (Fig.~\ref{fig:radius_az}), with extended emission beyond the axisymmetric model at PA$\sim$170$^{\circ}$ (Fig.~\ref{fig:B7_cont_R0p5}). Orbital plane tidal disruptions have been suggested by \cite{Dupree2026}, and further ALMA observations close to maximum elongation would be of interest, as the shocked and ionised gas may have a radio signature.

\section{Conclusions}

The sub-millimetre images of Betelgeuse with resolutions of $\sim$7-20~mas show an optically-thick atmosphere of free-free emission, with a radius of $\sim$1.2~$R_\star$ at $\lambda$0.6~mm. Mean temperatures of $\sim$2300K are consistent with models of the temperature inversion around this radius. At least two compact hot regions of emission are seen, including one to the north-east of the disk centre with a temperature $\sim$800K above the surrounding gas. We interpret these as localised energy input from underlying convection cells, with up to 1\% of the stellar luminosity at this wavelength. Comparison with an observation taken 7 years previously shows little change in this structure, indicating that they are relatively persistent compared with turbulent convection models.

The limb of the photosphere at these wavelengths shows significant deviations from symmetry, with corrugations of $\pm$6\% of the mean stellar radius. This may also be attributed to the effect of the unstable underlying stellar convection and, although these corrugations appear to have changed over 7 years, they are confined to only the southeast region of the star. 

Extended optically-thin free-free emission is seen from the atmosphere out to $\sim$2.2~$R_\star$. Multiple lines of SiO, CO, and their isotopologues are detected with transition energy levels $E_{low}$, up to 3500~K in emission, and up to 13800~K in absorption against the photosphere. Apart from maser spots in the $J$=5-4 lines, most molecular lines have a similar clumpy distribution in the MOLsphere of 1.3$-$3~$R_\star$, unlike the more symmetric extended free-free emission. The line emission appears to have changed significantly  compared with the previous observation, and the kinematics does not show clear evidence of rotation around a NE-SW axis. The hot continuum regions appear to be associated with broader blue-shifted absorption lines, suggesting the presence of upwelling gas.

\begin{acknowledgements}

This paper makes use of the following ALMA
data: ADS/JAO.ALMA\#2022.A.00026.S and ADS/JAO.ALMA\#2015.1.00206.S. ALMA is a partnership of ESO (representing its member states), NSF (USA) and NINS (Japan), together with NRC
(Canada), MOST and ASIAA (Taiwan), and KASI (Republic of Korea), in cooperation with the Republic of Chile. The Joint ALMA Observatory is operated by ESO, AUI/NRAO and NAOJ. LDM acknowledges support from award AST-2107681 from the National Science Foundation.
Support for GMH was provided by grant HST-GO-16256.001-A provided by Space Telescope Science Institute, which is operated by the Association of Universities for Research in Astronomy, Incorporated, under NASA contract NAS5-26555.
EOG acknowledges Tom Ray (DIAS) for the use of DIAS's computer facilities. We thank the staff at ALMA for their support, and in particular for routinely operating the array in the longest baseline configuration. 
We also thank the referee for extensive and insightful comments and suggestions. 

\end{acknowledgements}

\bibliography{my_biblio}{}
\bibliographystyle{aasjournal}

\begin{appendix}
\section{Observation simulations and imaging comparison}
\label{sec:appendix_sim}

We have run simulations of the observations to look for potential imaging artifacts due to the limited {\it uv} coverage in the ALMA long-baseline datasets and the enhanced weighting of long baselines when using the super-uniform option in image reconstruction. The thermal S/N in Band 8, derived from the image peak divided by the rms of the image off-source (clear of nearby imaging artifacts), is $\sim$26/0.23=113, when using super-uniform weighting (Fig.~\ref{fig:contin_images}). We wished to confirm that the image restoration was not generating generate additional artifacts, particularly when when a bright source is present.

\begin{figure}
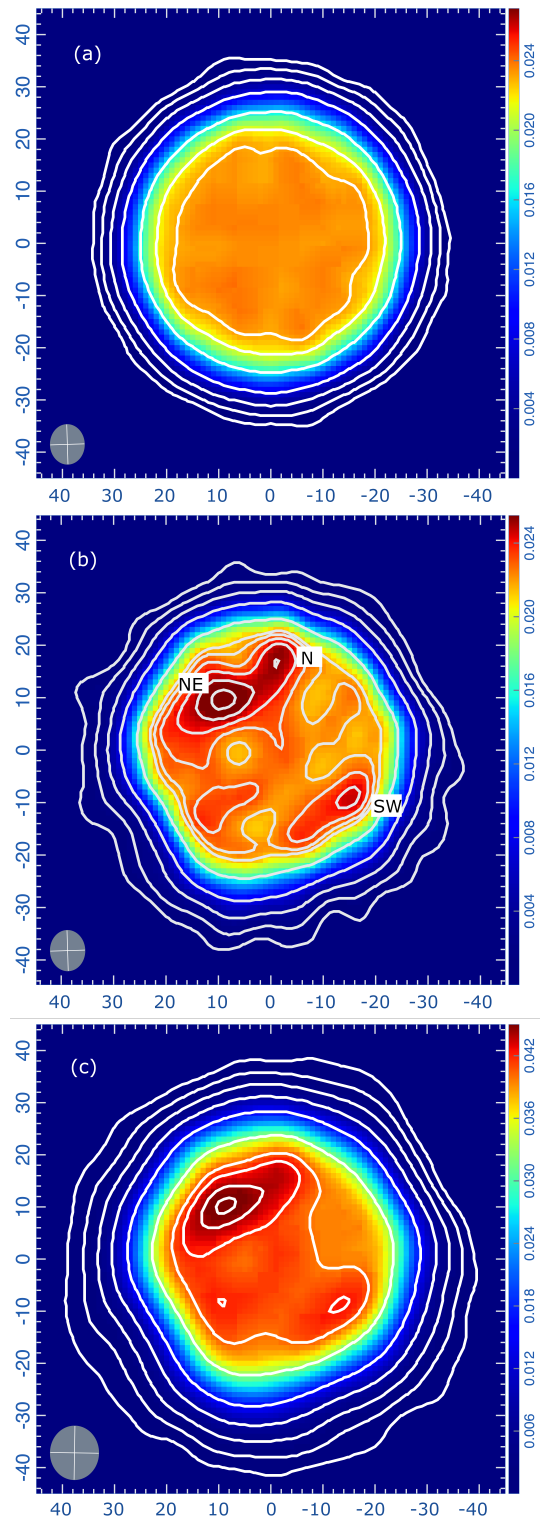

\begin{center}
 	\includegraphics[width=7.cm,angle=0]{FigA1a_B8_model.pdf}
 	\includegraphics[width=7.cm,angle=0]{FigA1b_B8_SUR.pdf}
 	\includegraphics[width=7.cm,angle=0]{FigA1c_B8_R0p5.pdf}
     \caption{Simulation vs. observations. (a) Simulated Band 8 observation of the axisymmetric smooth disk model described in Appendix~\ref{sec:appendix_model}, using super-uniform weighting in image reconstruction. (b) Real Band 8 observations with the same super-uniform weighting, The NE and SW hot regions are marked. Also marked 'N' is the additional point source included in the $uv$ fitting (Table~\ref{tab:uv_fits}). (c) Same Band 8 observations but with image reconstruction using Briggs weighting (R=0.5). Contours levels: 1,2,4,8,16,21,23,24,25,26~mJy/beam (5$\sigma$=1.2~mJy/beam) in (a) and (b), and 1,2,4,8,16,32,40,42,44,45~mJy/beam (5$\sigma$=0.85~mJy/beam) in (c). Offsets in mas, field of view 90~mas, and restoring beams lower left.
    }
	\label{fig:B8_model}
    \end{center}
\end{figure}

Simulations of the Band 7 and 8 data were performed by subtracting off the CLEAN component model of the star from the observed dataset, and then adding visbilities from the analytic axisymmetric model described in Appendix~\ref{sec:appendix_model}. This was done using the \textsc{aips} \texttt{UVSUB} task, and ensured the baseline coverage of the simulation was the same as the actual observation. It was then imaging using \textsc{casa} in the same way. Fig.~\ref{fig:B8_model} compares the reimaged Band 8 model with the actual data using the super-uniform weighting function. In the model image, the rms in the resolved bright photosphere region of the central disk is 0.25~mJy, compared with 0.24~mJy off the source. This close agreement suggests that the contribution from imaging artifacts is small. By comparison, the same region in the real data has an rms of $\sim$1.7mJy due to real structure. We also tested image reconstruction using Briggs weighting with robustness parameter 0.5 (lower panel); the resulting lower resolution image appears fully consistent with the super-uniform weighting.  As well as confirming the significance of the NE and SW peaks (Sect.~\ref{sec:surface_structure}), it also suggests that additional structure in the centre of the disk between the peaks may be significant, although confirmation must await deeper, high-resolution observations with good {\it uv} coverage.

The simulation in Fig.~\ref{fig:B8_model} also shows that the image reconstruction gives a fully axisymmetric structure - significantly different from the actual star, which shows notable radial corrugations in radius in the SE region (discussed in Sect.~\ref{sec:radial_structure}). 

\section{Semi-empirical thermodynamic radio continuum model}
\label{sec:appendix_model}

To understand the ALMA Band 7 and 8 observed radial intensity profiles shown in Fig.~\ref{fig:B78_spindex}
it is necessary to consider all the emitting material along the rays that traverse the atmosphere that can emit at these frequencies. In order to do so one needs to account for
the distribution of material that emits free-free continua at longer wavelengths and
this can be achieved by constructing a spherical spatially-resolved 
semi-empirical thermodynamic model (SEM) for the extended atmosphere and wind of Betelgeuse, 
building upon interferometic azimuthally-averaged real visibilities.

The SEM used to compare with the Band 7 and 8 data was constructed following the methodology described in detail by \cite{Harper2001} (hereafter HBL2001). 
The 1D spherical model describes the radial distribution of hydrogen density ($n_{\rm H}\>{\rm cm}^{-3}$), electron density ($n_e\>{\rm cm}^{-3}$) and electron (=gas) temperature (K).  Hydrogen
is predominantly neutral under both the lukewarm chromospheric and photospheric
conditions, and the electrons are from ultraviolet photoionization of abundant low First Ionization Potential ($< 13.6$\,eV) elements (i.e. Si, Mg, Fe) with
contributions from C and S in the warmer and lower density regions.

Molecular sub-mm opacity sources are not expected to be important throughout most of the SEM. They are not included here because of the absence of suitable photo-thermo-chemical molecular formation models, which are known to be important for the outer part of the SEM \citep{Beck1992}.
The presence of CO and SiO molecules does not directly affect the sub-mm opacity
because they are not significant contributors compared to H$^-$. Photoionized Si contributes free electrons, so the presence of SiO would reduce silicon's electron contribution, but SiO formation will be inhibited to some degree by the strong chromospheric ultraviolet radiation field \citep{Clegg1983}.
The partial pressure of H$_2$ in the coolest layers of early-M spectral-type supergiant photospheres is small compared to H \citep{Glassgold1983},
and the MARCS Betelgeuse model has partial pressure ratio, or ratio of molecular to atomic hydrogen density, of n$_{H_2}$/n$_H$ < $2.0\times10^{-3}$. Since H$^-$ is the
dominant sub-mm radio opacity, H$_2^-$ opacity is not expected to be significant in the photosphere or chromosphere.

In the coldest temperature minimum/MOLsphere regions molecular formation is known to occur, but unless the H$_2$/H ratio approaches 10\% it will not significantly affect the radio opacity because the sub-mm continuum opacity per hydrogen nuclei for H$_2^-$ free-free is $\sim$0.24 of that of H$^-$ free-free \citep{John1988}.

The SEM present here is a complete revision of the HBL2001 version because we adopt an accurate $K$-band continuum angular diameter of
$42.61\pm 0.05$ \citep{Montarges2021}, and improved VLA radio
visibility data, including for the first time ALMA data. 
The data adopted are: VLA Bands L, C, X and U \citep{OGorman2015}, VLA
Bands K, Q, and ALMA Bands 3 and 4 from \cite{Matthews2026}, and the ALMA Bands 6, 7, and 8 presented here.
For completeness we adopt the distance of 172\,pc and a stellar mass of $17.5\>M_\odot$
\citep{Macleod2025}. We adopt the K-band continuum to Rosseland $\tau=2/3$ angular diameter ratio
of 1.030, from the detailed interferometric and photospheric model RSG analysis by \cite{Arroyo-Torres2013},
which gives a Betelgeuse Rosseland $\tau=2/3$ radius of 812\,$R_\odot$.
The photospheric component of the SEM is obtained from a bi-linear
interpolation of spherical MARCS models \citep{Gustafsson2008}
for $T_{eff}=3650\>$\,K and a surface gravity of
$g_\ast=0.32\>{\rm cm\>s}^{-2}$, reduced from the adopted $g(R_\star)=0.73\>{\rm cm\>s}^{-2}$,
to mimic
the increased density scale-heights suggested by the
presence of macroscopic turbulence \citep{Josselin_2007},
and also from 3D hydrodynamical photospheric simulations of higher gravity red giant stars \citep{Ludwig2012}.

The SEM is constructed by assuming analytic representations for the hydrogen and electron densities 
in the turbulently-extended chromosphere and wind acceleration region, which are tied to the upper-photosphere
of the MARCS model. The initial guess of $T_e(R)$ is taken from the observed 
run of wavelength-dependent brightness temperature $T_{\rm b}$ with effective angular radius.
The radiative transfer equation is then solved for a spectrum of VLA and ALMA frequencies with
rays that intersect both the
opaque disk and the extended atmosphere - using an adaptive depth scale and
standard shell-ray geometry \citep{Harper1994}, and opacity sources given in HBL2001.

The resulting specific intensities are then used to generate 
real visibilities which are compared to those observed. Starting from the outermost observed flux (L band: 20.5\,cm) and visibilities (C band: 6.1\,cm) the parameters describing the atmospheric
density and temperature are manually adjusted and visibilities re-computed to get reasonable fits. Then the next inner visibility curve is optimized, i.e Band X (3.5 cm),
and so on until Band 8 is fit. The VLA and other ALMA radio data are non-contemporaneous and the star is known to be variable (see \citealt{OGorman2015}), and there are also
non-negligible absolute flux
uncertainties for each VLA and ALMA band. Therefore, emphasis was given to fitting
the visibility curves around the first two nulls to emphasize the spatial
information. The apparent size of the radio star is primarily controlled by the opacity through the density squared, i.e., $\propto n_e n_{\rm H}$, and the flux is then optimized by adjusting $T_e(R)$.
The temperature of the SEM is not unique, especially near the temperature minimum region (Bands 7 and 8) because they are not bracketed by constraints from ALMA Bands 9 and 10, which have not yet been observed.

Betelgeuse is known to harbour multiple dust reservoirs and these may affect
the sub-mm fluxes - particularly at the higher frequencies due to the opacity power law. The silicate shell at $\sim$1 arcsec radius was modeled by
\cite{Harper2001} and the dust flux contributions are shown in their Fig.~7.
The emission is diffuse and therefore not detectable in the present ALMA interferometric data. Amorphous alumina (Al$_2$O$_3$) has been suggested to exist close to the star by \cite{Verhoelst2006} in order to explain N-band spectra and near- and mid-ir interferometry.
\cite{Perrin2007} find an alumina optical depth of $\sim$1 near $\lambda$10$\mu m$ in a shell near 1.4$R_{\star}$
helps reproduce VLTI-MIDI interferometric data. This shell is in the same layers probed by our ALMA observations.
Taking the amorphous alumina optical constants of \cite{Suh2016} and
\cite{Begemann1997}, and using Mie scattering theory \citep{Bohren1983}
we compute the ratio of alumina absorption opacity at 0.85~mm and 10~$\mu m$ of $\sim$1.8$\times10^{-3}$.
If the alumina has the same turbulently-extended density-scale height as hydrogen,
then on the limb the optical depth will be <0.02. Since the dust temperature is $\sim$~1900~K its presence would not be detectable in our data.
Note the dust-scattering polarimetry of \cite{Haubois2023},
whose modelling favors enstatite (MgSiO$_3$), probe dust structures at similar
radii which are optically thin, and will therefore also not have a signature in the ALMA bands.

\begin{figure}
\begin{center}
 	\includegraphics[width=9.cm,angle=0]{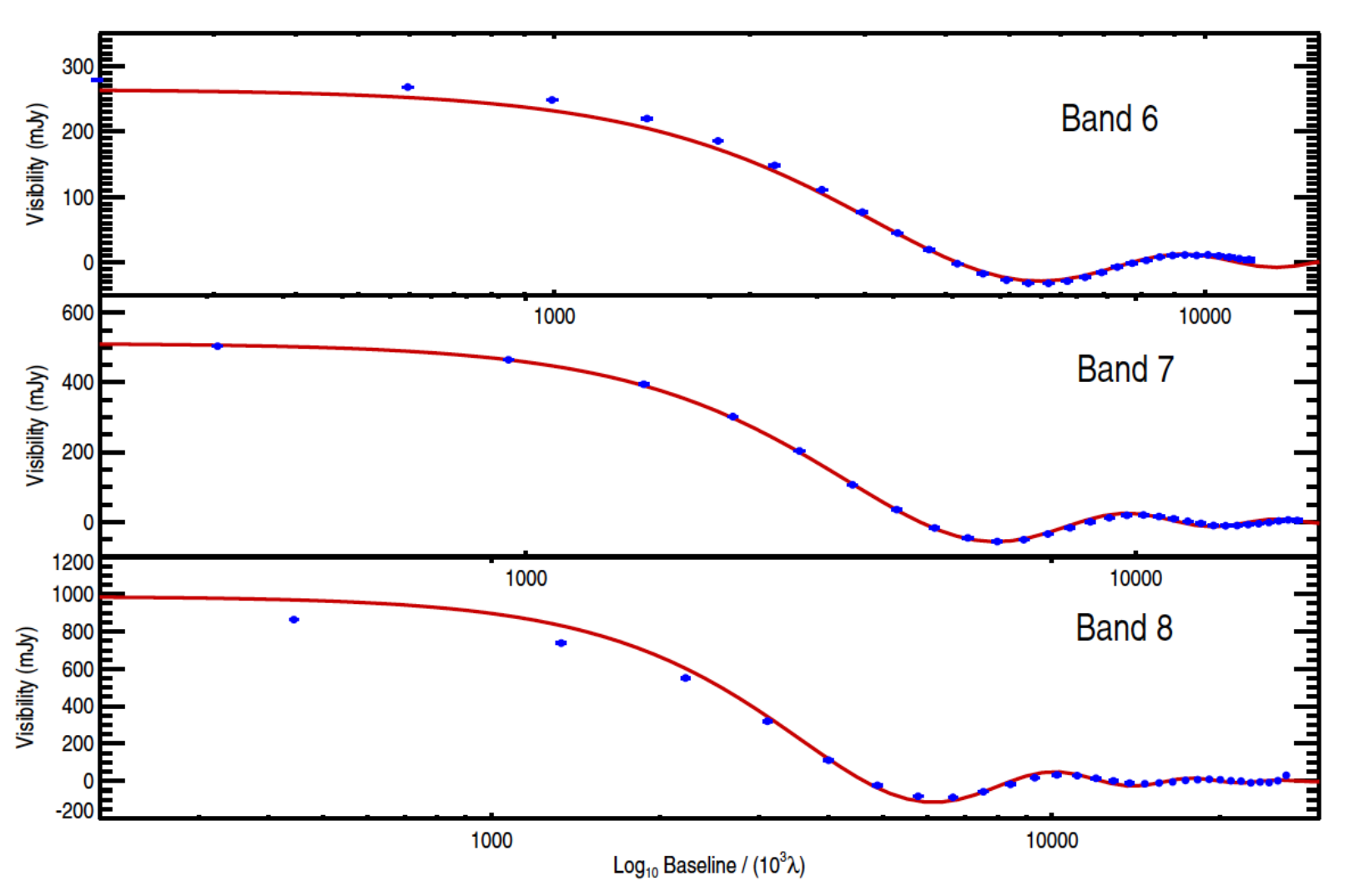}
     \caption{The computed and observed real visibilities for Bands 6, 7, and 8. The red curves are 
computed from the new SEM, and the blue circles are the azimuthally-averaged observations - the error bars are too small to see. Given the absolute flux uncertainties, emphasis was given to fitting the observations near the first two nulls. 
    }
	\label{fig:model_uv}
    \end{center}
\end{figure}

For the SEM adopted here, the predicted and azimuthally-averaged observed real visibilities for Bands 6, 7 and 8 are shown in Fig.~\ref{fig:model_uv}. The radial distribution of temperature is shown in Fig.~\ref{fig:T_model}. The hydrogen density above the base of the chromospheric region at $R_{min}=1.144R_\star$ is described by
\begin{equation}
n_H\left(R\right)= n_{chrom}e^{-(R-R_{min})/H} + n_\ast\left(R_\star\over{R}\right)^2 
\left(0.998 -\left({R_\star\over{R}}\right)^\gamma\right)^{-\delta}
\end{equation}
where $n_{chrom}=1.34\times 10^{12}\>{\rm cm}^{-3}$, and the hydrogen density scale-height, $H$, is given by
\begin{equation}
H = {v^2_{turb}\over{2g_\ast}}\left({R_{min}\over{R_\star}}\right)^2 = 0.06 R_\star
\end{equation}
where ${\rm v}_{turb}=19.5\>{\rm km\>s}^{-1}$ (most probable velocity) is a parameter expected to be similar 
to values observed in low optical-depth emission lines, e.g. C~II] 2325\AA ~\citep{Carpenter2018}.  Note that molecular features have a smaller ${\rm v}_{turb}=14.0\>{\rm km\>s}^{-1}$, but the optimum model favours the higher value.
The adopted wind acceleration terms
are: $n_\ast=2.96\times 10^{9}\>{\rm cm}^{-3}$, $\gamma=0.45$ and $\delta=1.50$.

The ionization fraction, $n_e/n_{\rm H}$, is linearly interpolated in radius from $n_e/n_{\rm H}=2.5\times 10^{-5}$
in the cooler dense plasma at $R_{min}$ to 
$n_e/n_{\rm H}=5.0\times 10^{-4}$ in the warmer lower density region at $2R_\star$, and kept constant
farther out. 

For a helium abundance of $A({\rm He})=0.083$, if the wind velocity tends to
$9\>{\rm km\>s}^{-1}$ (the velocity of the slower and inner S1 circumstellar outflow: \citealt{Goldberg1975, Bernat1979, OGorman2012}), then, assuming a steady spherical outflow, the total
mass-loss rate would be $\dot{M}=3.8\times 10^{-6}\>M_\odot\>{\rm yr}^{-1}$.

In this SEM, the radial $\tau=1$ depths for Bands 6, 7 and 8 are $\simeq$ 1.18, 1.15, and 1.13 $\>R_\star$,
respectively, which shows that we are probing the  coolest layers yet in the radio. The minimum gas temperature of $\sim 1700$\,K is less than $T_{\rm b}$ for Bands 7 and 8 because $T_{\rm b}$ represents an average value which includes contributions from warmer plasma. The
$T_{\rm e}$-structure near the minimum should only be taken as indicative because 
the actual atmosphere is not axisymmetric (see Fig.~\ref{fig:radius_az}), and the lack of empirical and theoretical constraints in this region.
Higher frequency bands would allow us to probe closer to the optical photosphere and improve the current model. 

The deduced run of $n_{\rm H}(R)$, $n_e(R)$, and $T_e(R)$ presented here combined with observed HST far-ultraviolet spectra should enable future studies of SiO and CO formation within the temperature minimum and lukewarm chromosphere. Future progress can also be made by constructing models for different 
angular sectors to account for the observed non axisymmetry.

\newpage
\section{Integrated spectra}

\label{sec:appendix_spectra}

\begin{figure}[h]
\begin{center}
 	\includegraphics[width=9.1cm,angle=0]{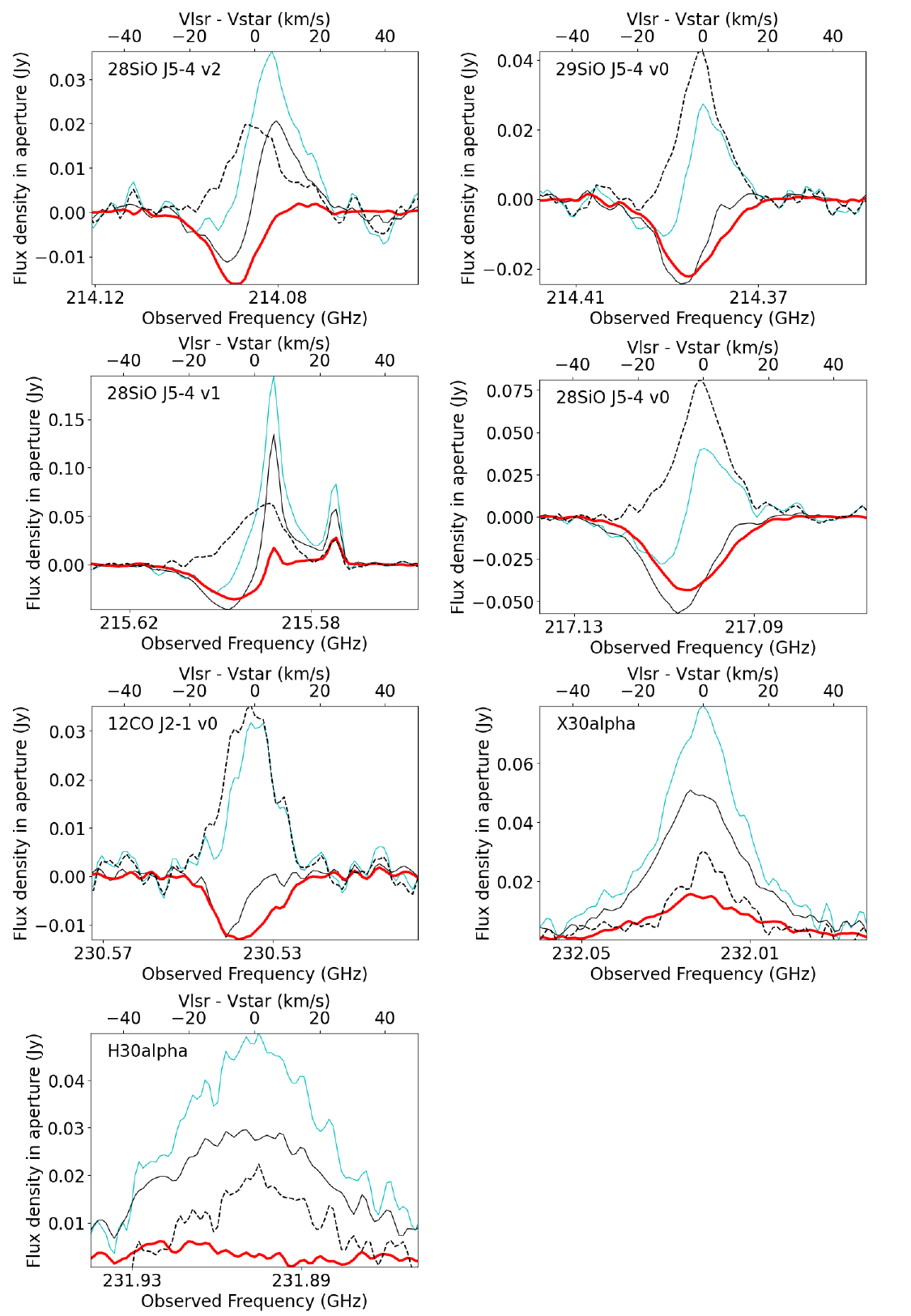}
     \caption{Integrated spectra in different apertures centred on the star in Band 6. Red lines show absorption against the photosphere, and back dashed lines show emission from the surrounding region - see text for details. Velocity scale covers $\pm$~50 km s$^{-1}$ with respect to 
     ${\rm v}_{\star}$.  }
    
	\label{fig:all_spectra1}
    \end{center}
\end{figure}

\begin{figure}
\begin{center}
    \includegraphics[width=9.1cm,angle=0]{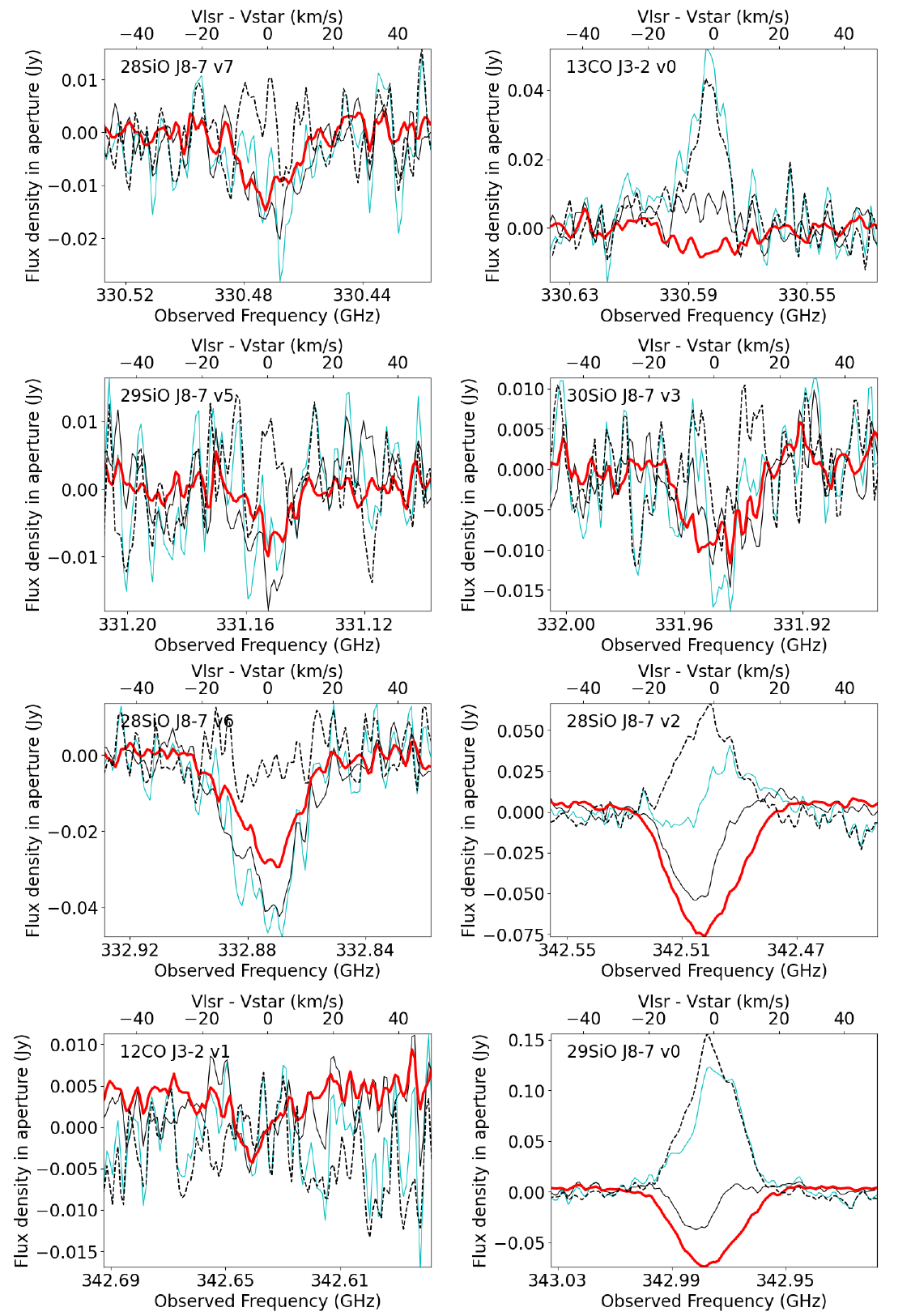}
     \caption{Integrated spectra in different apertures centred on the star in Band 7.}
    
	\label{fig:all_spectra2}
    \end{center}
\end{figure}

\begin{figure}
\begin{center}
    \includegraphics[width=9.1cm,angle=0,trim=0cm 4cm 0cm 0cm,clip]{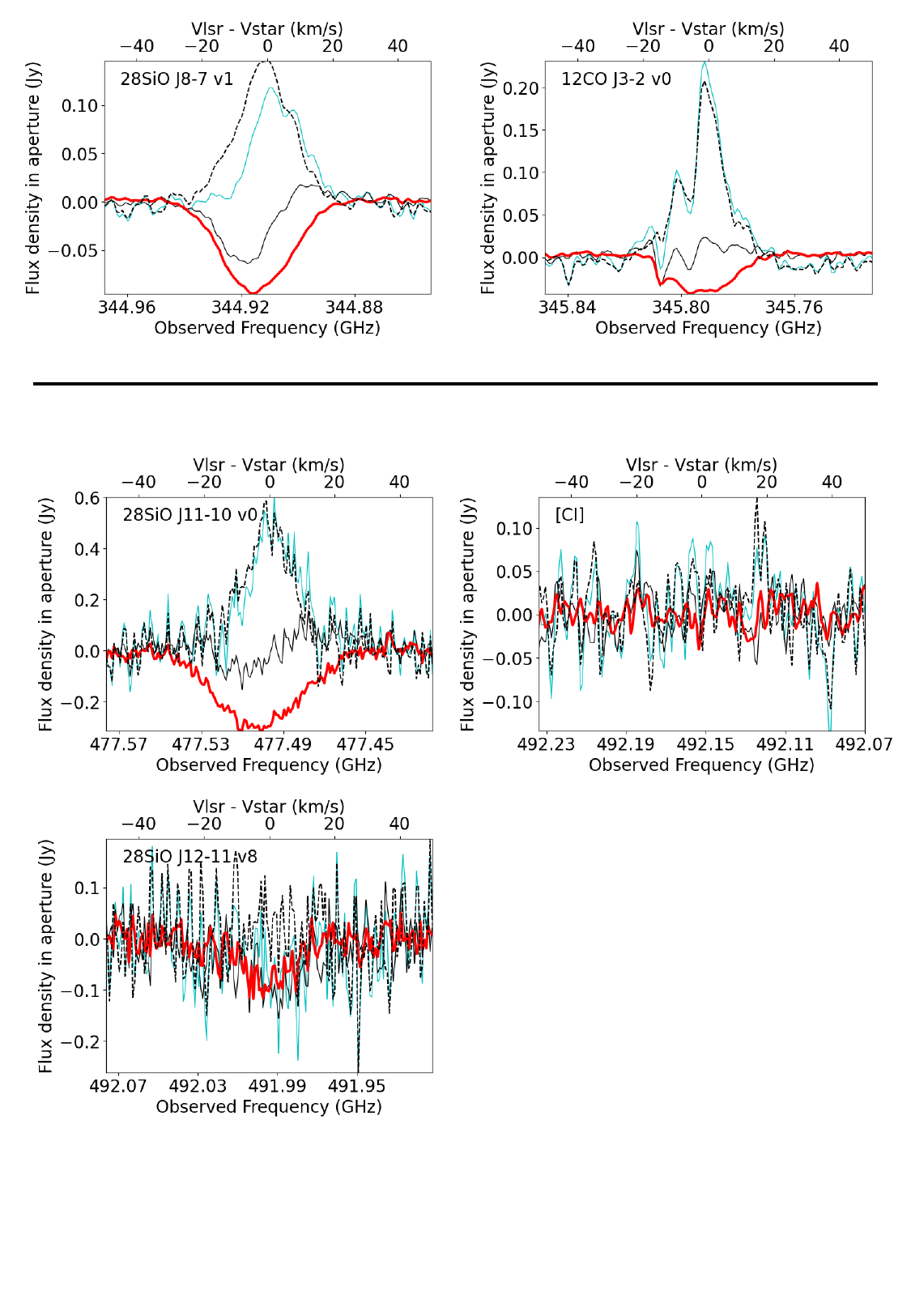}
     \caption{Integrated spectra in different apertures centred on the star in Band 7 (cont.) and 8.}
    
	\label{fig:all_spectra3}
    \end{center}
\end{figure}

The spectra of observed lines are given in Figs.\ref{fig:all_spectra1} - \ref{fig:all_spectra3}. Spectra have been smoothed using Hanning smoothing with a window of 5 channels. The line colours show the integrated flux in circular apertures centred on the star of radii 22~mas (red, covering the photosphere), 44~mas (black), 88~mas (cyan) and an annulus of radii 44-88~mas (black dashed, avoiding the photosphere). Spectral range is 100~km s$^{-1}$ centred on the catalogue rest frequency with a stellar velocity of ${\rm v}_{\mathrm{LSR}}$~=+4.9 km s$^{-1}$ .

\FloatBarrier

\section{Channel maps}
\label{sec:appendix_chanmap}

\begin{figure}[!h]
    \centering
    \noindent
    \makebox[\textwidth][l]{%
        \begin{minipage}[l]{0.65\textwidth}
            \centering
            \includegraphics[width=9cm,angle=-90,trim=0cm 2cm 1cm 0cm,clip]{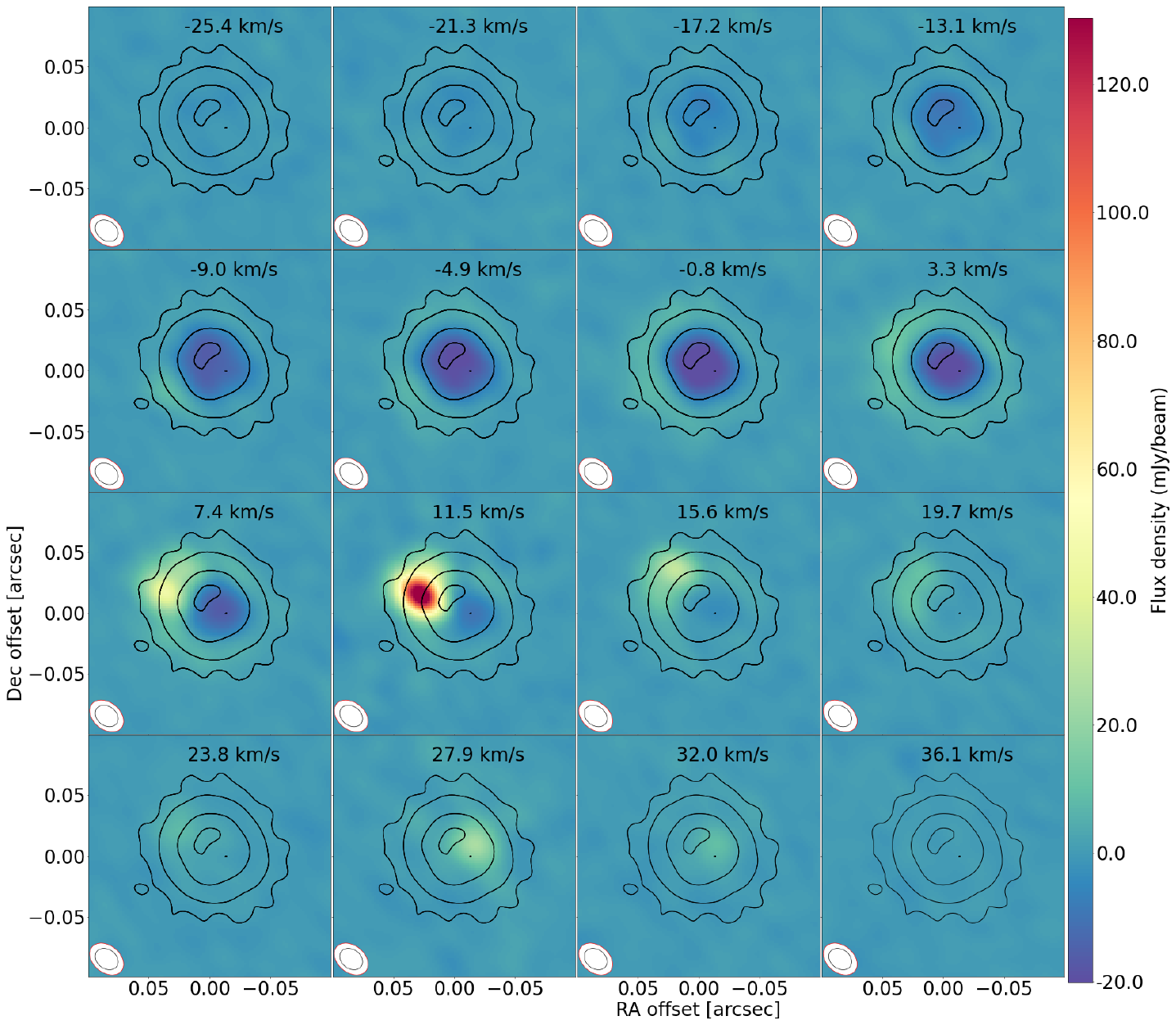}
        \end{minipage}
        \hspace{0.0\textwidth}
        \begin{minipage}[c]{0.25\textwidth}
            \caption{Channel maps of SiO $v$=1 $J$=5-4 (Band 6). This shows the bright maser spot at $\sim$11~km s$^{-1}$. The SiO $v$=2 $J$=5-4 distribution is similar, although the maser spot is less prominent. The contours of Band 6 continuum emission are superimposed, and the restoring beams in continuum and line are shown lower left (beam for the line data is the larger of the two).}
            
            \label{fig:SiO_B6_channel}
        \end{minipage}%
    }
\end{figure}

\begin{figure}[!h]
    \centering
    \noindent
    \makebox[\textwidth][l]{%
        \begin{minipage}[l]{0.65\textwidth}
            \centering
            \includegraphics[width=9cm,angle=-90,trim=0cm 2cm 1cm 0cm,clip]{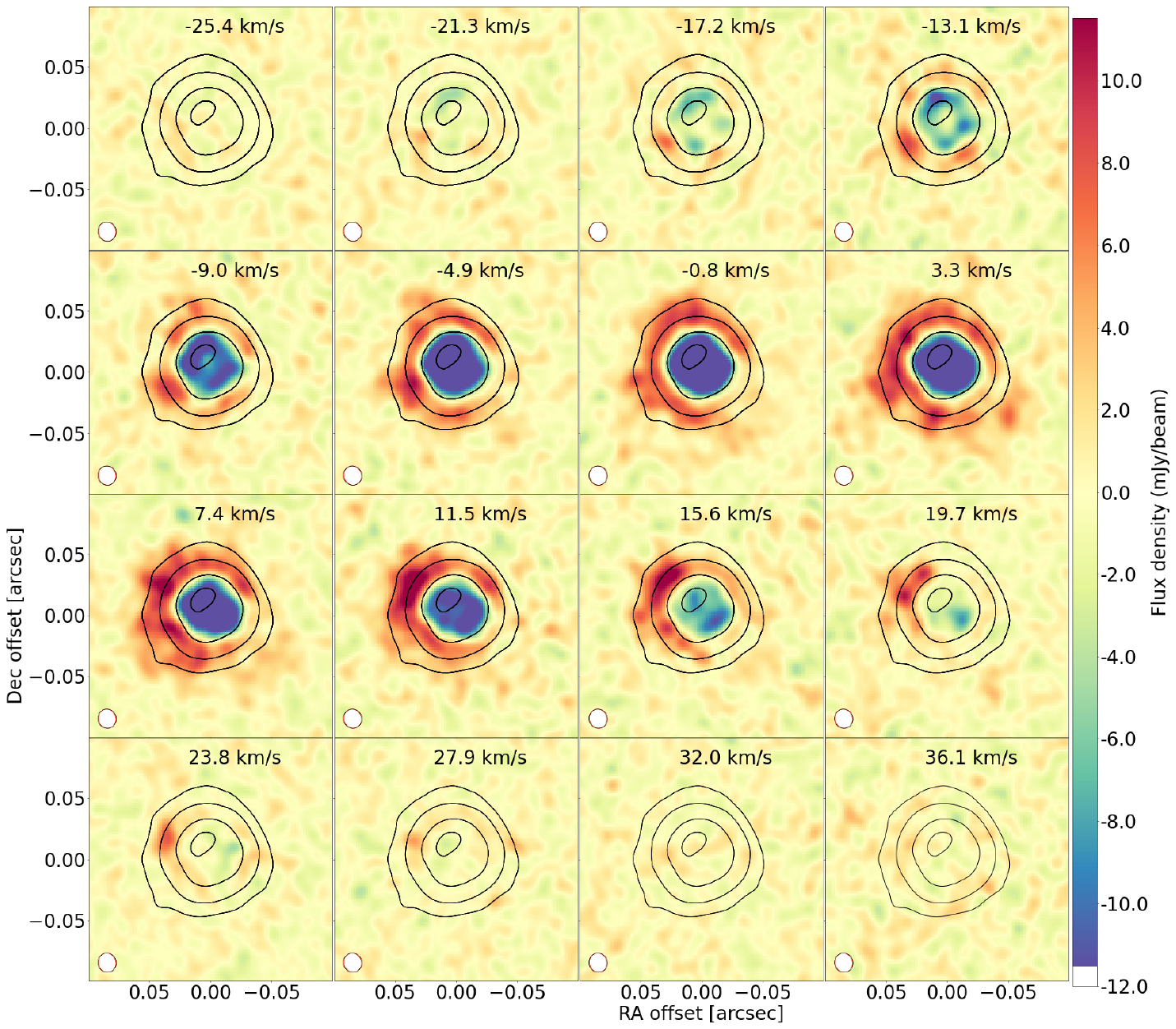}
        \end{minipage}%
        \hspace{0.0\textwidth}%
        \begin{minipage}[c]{0.25\textwidth}
            \caption{Channel maps of SiO $v$=1 $J$=8-7 (Band 7). The contours of Band 7 continuum emission are superimposed, and the beams are shown lower left.}
            
            \label{fig:SiO_B7_channel}
        \end{minipage}%
    }
\end{figure}

\begin{figure}[!h]
    \centering
    \noindent
    \makebox[\textwidth][l]{%
        \begin{minipage}[l]{0.65\textwidth}
            \centering
            \includegraphics[width=9cm,angle=-90,trim=0cm 2cm 1cm 0cm,clip]{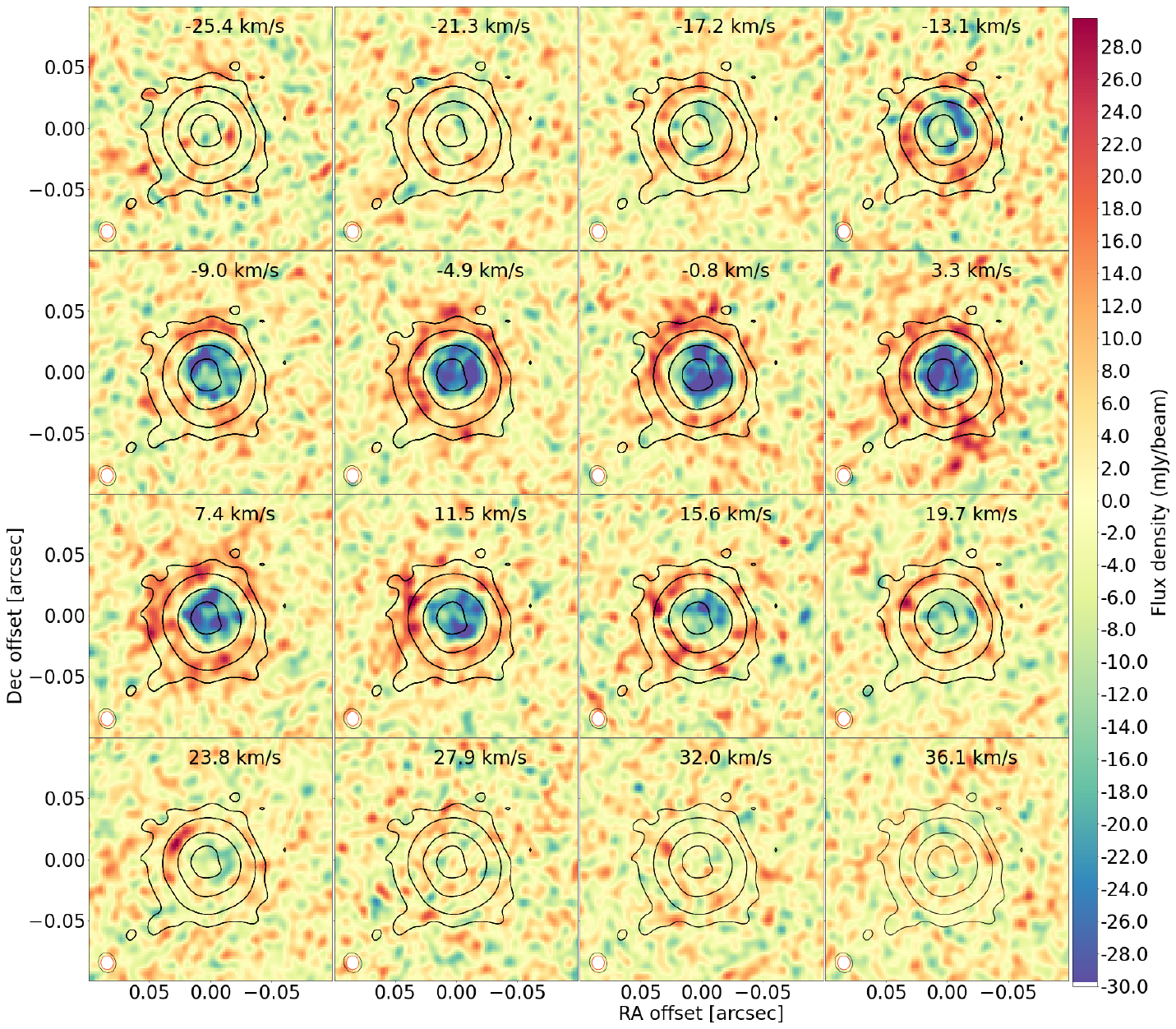}
        \end{minipage}%
        \hspace{0.0\textwidth}%
        \begin{minipage}[c]{0.25\textwidth}
            \caption{Channel maps of SiO $v$=0 $J$=11-10 (Band 8). The contours of Band 8 continuum emission are superimposed, and the beams are shown lower left.}
            
            \label{fig:SiO_B8_channel}
        \end{minipage}%
    }
\end{figure}

\begin{figure}[!h]
    \centering
    \noindent
    \makebox[\textwidth][l]{%
        \begin{minipage}[l]{0.65\textwidth}
            \centering
            \includegraphics[width=9cm,angle=-90,trim=0cm 2cm 1cm 0cm,clip]{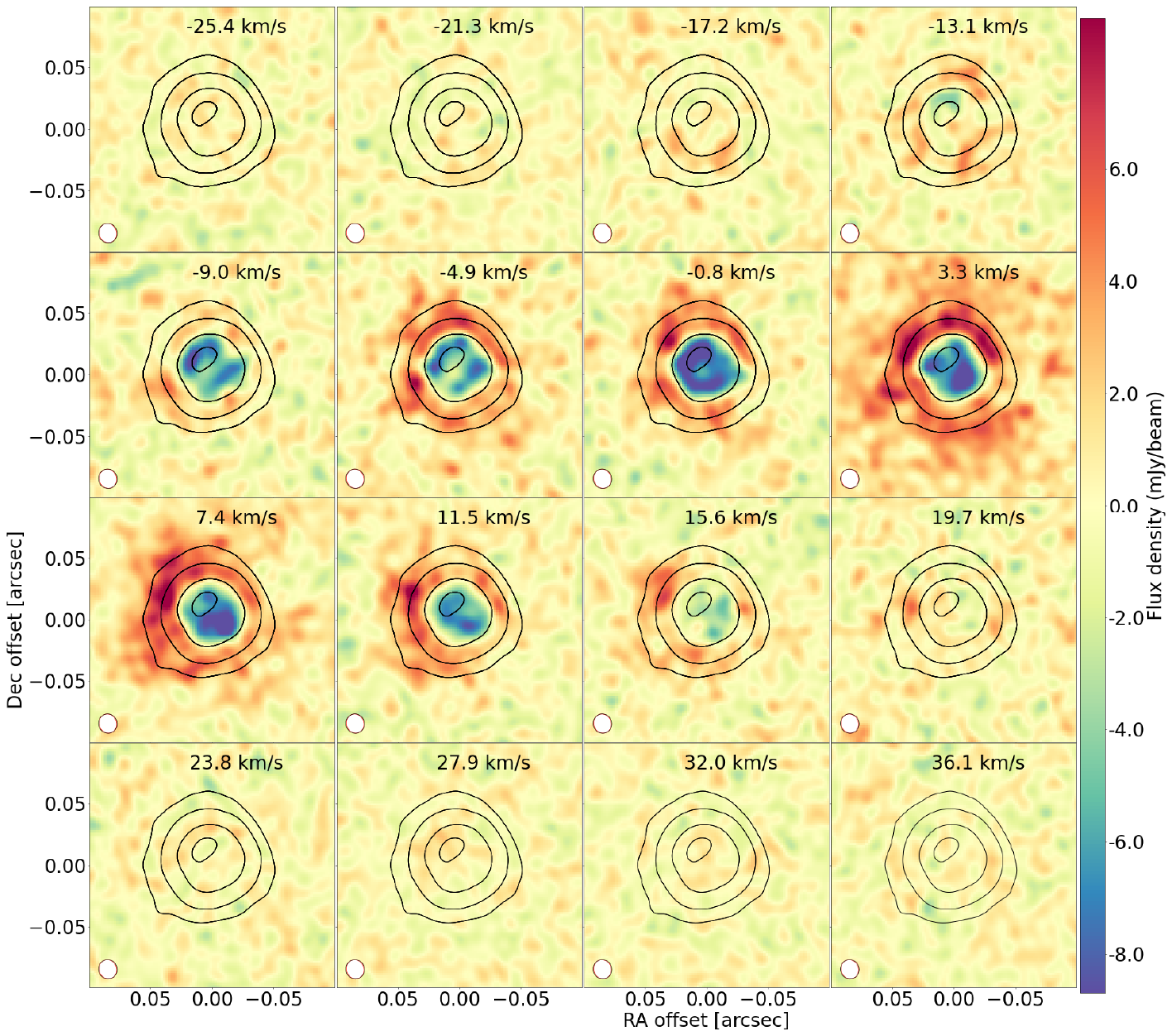}
        \end{minipage}%
        \hspace{0.0\textwidth}%
        \begin{minipage}[c]{0.25\textwidth}
            \caption{Channel maps of CO $J$=3-2 $v$=0 (Band 7). The contours of continuum emission are superimposed, and the beams are shown lower left.}
            
            \label{fig:CO_channel}
        \end{minipage}%
    }
\end{figure}

Examples of channel maps are given in Fig.~\ref{fig:SiO_B6_channel} (Band 6, $^{28}$SiO $v$=1 $J$=5-4), \ref{fig:SiO_B7_channel} (Band 7, $^{28}$SiO $v$=1 $J$=8-7), \ref{fig:SiO_B8_channel} (Band 8, $^{28}$SiO $v$=0 $J$=11-10) and \ref{fig:CO_channel} (Band 7, $^{12}$CO $v$=0 $J$=3-2). Fig.~\ref{fig:SiO_B6_channel} shows the bright maser spot, which is not seen in the higher $J$ transitions. Distributions of Band 7 and 8 SiO $v$=0, 1 and 2 and CO lines are similar to those shown in Fig.~\ref{fig:SiO_B7_channel}, \ref{fig:SiO_B8_channel} and \ref{fig:CO_channel}. Higher $v$ transitions are mostly seen only in absorption against the stellar continuum. 

\twocolumn
\newpage

\FloatBarrier 
\clearpage

\end{appendix}
\end{document}